\documentclass[12pt,a4papers,epsfig]{article}
\usepackage{a4wide}
\usepackage{amsmath}
\usepackage{amssymb}
\usepackage{amsxtra}
\usepackage{epsfig}
\usepackage{nameref}
\usepackage{color}
\usepackage{subfigure}
\usepackage{exscale}
\usepackage{float}
\usepackage{bbm}
\usepackage[numbers,sort&compress]{natbib}

\usepackage{color}
\usepackage{graphicx}

\newcommand{\RP}{{\mathbb{R}P}}

\newcommand{\Z}{{\mathbb{Z}}}

\makeatletter
\@addtoreset{equation}{section}
\makeatother

\title{Finite-Volume Energy Spectrum, Fractionalized Strings, and Low-Energy 
Effective Field Theory for the Quantum Dimer Model on the Square Lattice}

\author{D.\ Banerjee$^{a,b}$, M.\ B\"ogli$^{a,c}$, C.\ P.\ Hofmann$^d$, F.-J.\
Jiang$^e$, P.\ Widmer$^a$,\\
and U.-J.\ Wiese$^a$ \\ \\
$^a$ Albert Einstein Center for Fundamental Physics \\ 
Institute for Theoretical Physics, Bern University \\
Sidlerstrasse 5, CH-3012 Bern, Switzerland \\ \\
$^b$ NIC, DESY, Platenenallee 6, 15738 Zeuthen, Germany \\ \\
$^c$ Department of Physics, Chung-Yuan Christian University (CYCU), \\
Chung-Li 32023, Taiwan \\ \\
$^d$ Facultad de Ciencias, Universidad de Colima \\
Colima C.P.\ 28045, Mexico \\ \\
$^e$ Department of Physics, National Taiwan Normal University \\
88, Sec.\ 4, Ting-Chou Rd., Taipei 116, Taiwan}

\begin{document} 

\maketitle

\begin{abstract} \normalsize

We present detailed analytic calculations of finite-volume energy spectra, 
mean field theory, as well as a systematic low-energy effective field theory
for the square lattice quantum dimer model. The analytic considerations
explain why a string connecting two external static charges in the confining
columnar phase fractionalizes into eight distinct strands with electric flux 
$\frac{1}{4}$. An emergent approximate spontaneously broken $SO(2)$ symmetry 
gives rise to a pseudo-Goldstone boson. Remarkably, this soft phonon-like 
excitation, which is massless at the Rokhsar-Kivelson (RK) point, exists far 
beyond this point. The Goldstone physics is captured by a systematic low-energy 
effective field theory. We determine its low-energy parameters by matching the 
analytic effective field theory with exact diagonalization results and Monte
Carlo data. This confirms that the model exists in the columnar (and not in a
plaquette or mixed) phase all the way to the RK point.

\end{abstract}

%\pacs{74.20.Mn, 73.43.Nq, 75.10.Jm, 78.20.Bh}

\section{Introduction}

Despite the extensive work on high-temperature superconductivity during the
past decades since their discovery \citep{BM86}, understanding the mechanism
of electron or hole pairing still represents a major unsolved problem in
condensed matter physics. One of the various proposed scenarios is related to
the quantum dimer model that was introduced by Rokhsar and Kivelson in
Ref.~\citep{Rok88}. It represents a simple realization of the resonating
valence bond (RVB) state, proposed by Anderson in his pioneering paper
\citep{And87}, and provides a possible route towards understanding
high-temperature superconductivity. Quantum dimer models have attracted a lot
of attention over the years, as they are also relevant beyond high-temperature
superconductivity, e.g., in connection with deconfined quantum criticality or
topological order. Unraveling the phase structure of both the classical and 
the quantum dimer model has been the subject of many publications
\citep{Sac89,Lev90,Leu96,Moe02,Hen04,Ale05,Ale06,Cha10,Can10,Alb10,Tan11,Lam13,
Syl05}.

These studies include dimer models on both bipartite and non-bipartite
lattices, which are defined in spatial dimensions $d \ge 2$. Quite
surprisingly, even in the case of the simple square lattice the question of
which phases are realized as a function of the Rokhsar-Kivelson (RK) parameter
$\lambda$ has been controversial. This may even be more surprising in view of
the fact that Monte Carlo simulations of quantum dimer models on the square
lattice are not affected by the sign problem. While some authors claimed that
a plaquette phase arises from a columnar phase in a first order phase
transition around $\lambda \approx 0.6$ \citep{Syl06}, other studies found
evidence for a mixed phase for $\lambda \gtrsim 0$, exhibiting features of both
the columnar and the plaquette phase \citep{Ral08}.

In a recent study \citep{BBHJWW14}, using quantum Monte Carlo applied to dual height
variables as well as exact diagonalization, we have challenged these various views.
In particular, we pointed out that there is no evidence for a plaquette or
mixed phase in the square lattice quantum dimer model --- rather the columnar
phase extends all the way to the RK point at $\lambda = 1$. Moreover, we showed
that two external static charges $\pm 2$ are confined by an electric flux
string that fractionalizes into eight strands carrying fractionalized flux
$\frac{1}{4}$. Inside these strands, which represent interfaces separating
different columnar orders, we found plaquette phase. However, the plaquette
phase only exists inside the strands and not in the bulk. Finally, as a
consequence of an approximate emergent $SO(2)$ symmetry, we found evidence for
a soft pseudo-Goldstone boson that exists in the parameter regime
$0 \lesssim \lambda < 1$, i.e., even far beyond the RK point.

In the present article we complement our previous Monte Carlo and exact
diagonalzation results with detailed analytic calculations of finite-volume
energy spectra, mean field theory, as well as a systematic low-energy
effective field theory for the pseudo-Goldstone boson. Overall, we consolidate
our previous findings that contradict the earlier views on the phase structure
of the square lattice quantum dimer model.

The paper is organized as follows. In section \ref{ModelObservables} we define
the quantum dimer model and discuss its symmetries on the square lattice. We
then introduce height variables on the dual lattice. On the one hand, these
allow us to define order parameters that distinguish the various candidate
phases. On the other hand, the dual height variables are the basic degrees of
freedom on which our Monte Carlo simulations operate. We also perform a
systematic mean field analysis of the quantum dimer model. In Section
\ref{FVES} we investigate the finite-volume energy spectrum as a diagnostic of
the phase structure. Section \ref{LEEFT} is dedicated to the low-energy
effective field theory for the soft pseudo-Goldstone mode and the
corresponding rotor spectrum. In section \ref{ExactDiag} we present
exact diagonalization results and use them to estimate some low-energy
parameters of the effective field theory. In Section \ref{MC} we present new
Monte Carlo data for the confining strings in the columnar phase. Finally, in
section \ref{conclusions} we present our conclusions. Appendix A summarizes
the symmetry properties of the relevant order parameters.

\section{Model and Observables}
\label{ModelObservables}
In this section, we consider the quantum dimer model and discuss its
symmetries on the square lattice. We then define height variables on the dual
lattice, which are the basic degrees of freedom in our Monte Carlo
simulations. They also serve to construct order parameters that signal which
phase is realized. Finally, we perform a systematic mean-field analysis with the
intention to gain qualitative insight into this question.

\subsection{Model}
\label{Model}

\begin{figure}
\begin{center}
\includegraphics[scale=1.0]{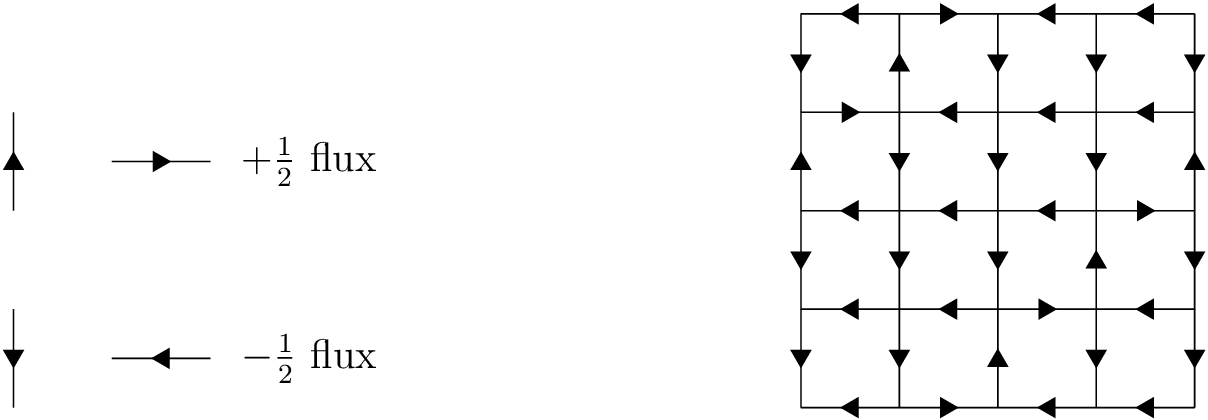}
\end{center}
\caption{Definition of flux states (left) and typical flux configuration on
the square lattice U(1) quantum link model (right).}
\label{fluxStatesConfiguration}
\end{figure}

The Hamiltonian of the quantum dimer model coincides with the Hamiltonian of
the $(2+1)$-d $U(1)$ quantum link model \citep{Hor81,Orl90,Cha97}. However, the
corresponding Gauss law is realized differently. The Hamiltonian of both the
$U(1)$ quantum link model and the quantum dimer model takes the form
\begin{equation}
\label{Hamiltonian}
H = - J \sum_{\Box} \left[U_\Box + U_\Box^\dagger - \lambda
(U_\Box + U_\Box^\dagger)^2\right].
\end{equation}
In the above Hamiltonian, the quantity
$U_\Box = U_{wx} U_{xy} U_{zy}^\dagger U_{wz}^\dagger$ represents a plaquette
operator expressed in terms of quantum links $U_{xy}$ that connect the
nearest-neighbor sites $x$ and $y$ on the square lattice. A $U(1)$ quantum link
$U_{xy} = S_{xy}^+$ is a raising operator of the electric flux
$E_{xy} = S_{xy}^3$, which is built from a quantum spin $\frac{1}{2}$ associated
with the link $xy$.  In the $U(1)$ quantum link model, each link has two
possible states characterized by electric flux $\pm \frac{1}{2}$, represented
pictorially by arrows as shown in Fig.~\ref{fluxStatesConfiguration}. A
typical flux configuration of the $U(1)$ quantum link model is depicted in the
same figure.

\begin{figure}
\begin{center}
\includegraphics[scale=1.0]{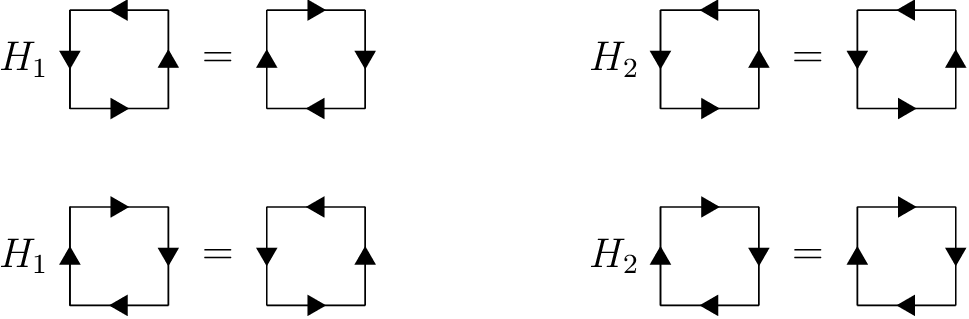}
\end{center}
\caption{The results of applying the Hamiltonian of Eq.~(\ref{Hamiltonian}) to
some plaquette flux states. Here $H_1$ and $H_{2}$ represent the terms
in Eq.~(\ref{Hamiltonian}) proportional to $J$ and $J \lambda$, respectively.
When the Hamiltonian acts on other plaquette configurations (which are not
shown explicitly) the result vanishes.}
\label{act2}
\end{figure}

\begin{figure}
\begin{center}
\includegraphics[scale=1.0]{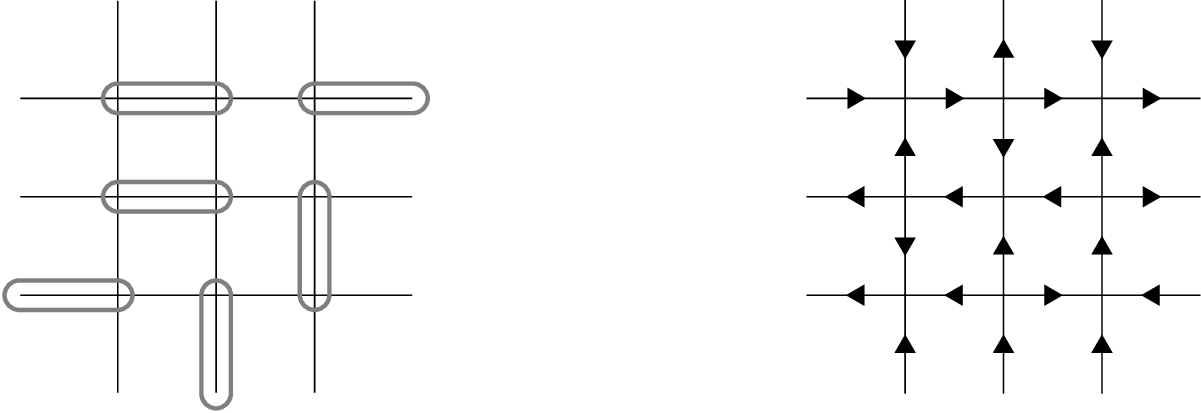}
\end{center}
\caption{Mapping between dimer and flux configurations.}
\label{TranslationDimerFlux}
\end{figure}

Applying the Hamiltonian of Eq.~(\ref{Hamiltonian}) to a plaquette flux state
leads to the results shown in Fig.~\ref{act2}. In summary, the first
contribution to the Hamiltonian (\ref{Hamiltonian}), proportional to the
parameter $J$, flips a loop of flux that winds around a plaquette. Flux states
that do not correspond to closed flux loops are referred to as non-flippable
plaquettes, which are annihilated by the Hamiltonian. On the other hand, the
second contribution to the Hamiltonian (\ref{Hamiltonian}), proportional to the
RK parameter $\lambda$, counts the plaquettes that are flippable. 
Notice that the configurations of the
square lattice quantum dimer model are characterized in terms of  variables
$D_{xy} \in\{0,1\}$ which signal whether a dimer is present or absent on the
link that connects two neighboring sites $x$ and $y$. In addition, the
electric flux variables $E_{xy}$ can be expressed through the dimer variables
$D_{xy}$ as
\begin{equation}
E_{xy} = (-1)^{x_1+x_2}(D_{xy}-\frac{1}{2}) .
\end{equation}
This mapping between a dimer and a flux configuration of the quantum dimer
model is illustrated in Fig.~\ref{TranslationDimerFlux}.

Notice that, in the $U(1)$ quantum link model, the physical state
$|\psi\rangle$ satisfies
\begin{equation}
\label{GaussQLM}
G_x |\psi \rangle = 0,
\end{equation}
where the quantity
\begin{equation} 
\label{generators}
G_x = \sum_{i}(E_{x,x+\hat{i}}-E_{x-\hat{i},x} )
\end{equation}
commutes with the Hamiltonian and describes an infinitesimal $U(1)$ gauge
transformation. Here $\hat i$ is the unit-vector in the $i$-direction.
Eq.~(\ref{GaussQLM}) represents the Gauss law for the $U(1)$ quantum link
model. In the quantum dimer model, using the connection between the electric
flux and the dimer variables, one has
\begin{equation}
G_x = (-1)^{x_1+x_2} \sum_i (D_{x,x+\hat i} + D_{x-\hat i,x}) = (-1)^{x_1+x_2}.
\end{equation} 
In other words, the dimer covering constraint implies that the quantum dimer
model is characterized by background electric charges $\pm 1$ that are
arranged in a staggered pattern. Accordingly, physical states in the quantum
dimer model satisfy 
\begin{equation}
G_x |\Psi\rangle = (-1)^{x_1+x_2} |\Psi\rangle .
\end{equation} 

\subsection{Symmetries}
\label{Symmetries}

The quantum dimer model on the square lattice exhibits various symmetries. We
first have a continuous $U(1)$ gauge symmetry and a global $U(1)^2$ center
symmetry. The latter is associated with ``large" gauge transformations
\citep{tHo79}. The model also has various discrete global symmetries. These
include translations by one lattice vector followed by charge conjugation
($CT_x$ and $CT_y$), which are equivalent to ordinary translations of the
dimers $D_{xy}$. Note that charge conjugation changes the sign of all electric
flux variables. It is important to point out that, in contrast to the quantum
link model, the transformations $T_x, T_y$ and $C$ individually are not
symmetries of the quantum dimer model because they are explicitly violated by
the Gauss law. Furthermore we have 90 degrees rotations around a plaquette
corner ($O$), 90 degrees rotations around a plaquette center followed by
charge conjugation ($CO'$), and finally, reflections on the $x$- and $y$-axes
($R_x$ and $R_y$). Below we will construct order parameters that will help us
to determine which phases are realized in the square lattice quantum dimer
model. It is then crucial to know how these different order parameters
transform under the various symmetries (see Sec.~\ref{OPandPhases} and
Appendix \ref{appendixA}).

\subsection{Dual Height Variables}
\label{DualVariables}

In this subsection we introduce height variables that reside on the dual
lattice. This height representation of the quantum dimer model is essential in
our approach.  It allows us, on the one hand, to design cluster and Metropolis
algorithms that operate in the space of these height variables and, on the
other hand, to construct order parameters to unambiguously distinguish the
various phases. 

\begin{figure}
\begin{center}
\includegraphics[scale=1.0]{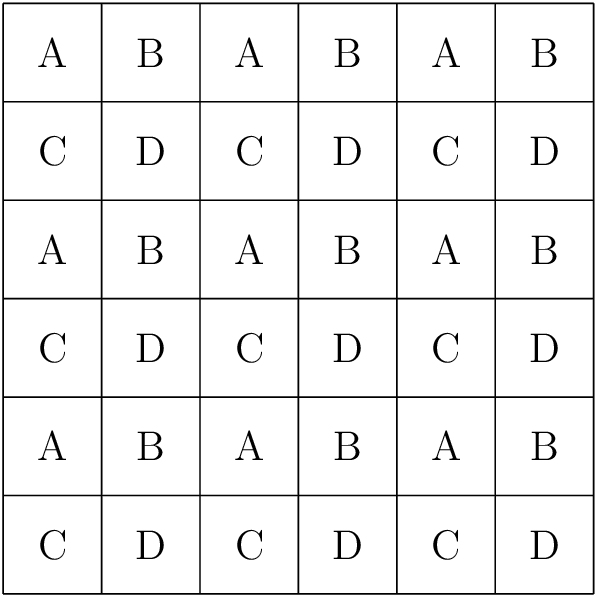}
\caption{The four dual sublattices $A, B, C, D$ used in the construction of
the height variables $h^{A,B,C,D}$.}
\label{DualHeightVariables}
\end{center}
\end{figure}

As illustrated in Fig.~\ref{DualHeightVariables}, in the case of the square
lattice quantum dimer model, we define four dual sublattices $A, B, C$, and
$D$, which consist of the points
\begin{equation}
\tilde{x} = (x_1+\frac{1}{2},x_2+\frac{1}{2}) .
\end{equation}
Each of the dual sublattices $X$ carries dual height variables $h^X$ that take
the values
\begin{equation}
h^{A,D}_{\widetilde x} = 0,1 , \qquad h^{B,C}_{\widetilde x} = \pm \frac{1}{2} .
\end{equation}

\begin{figure}
\begin{center}
\includegraphics[scale=1.0]{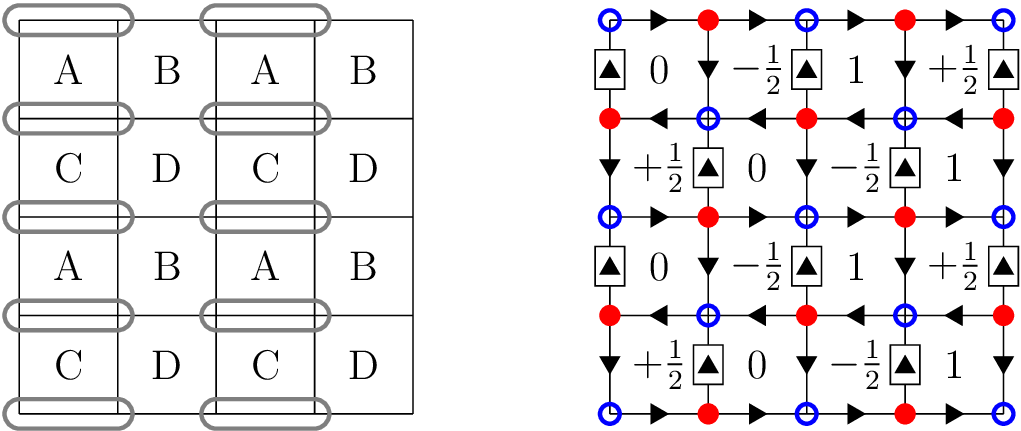}
\caption{The mapping between the dimer configuration and the corresponding
flux and height representation for a columnar quantum dimer configuration.}
\label{dimerConfigFluxHeight}
\end{center}
\end{figure}

When defining the height variables on the dual lattice, we will encounter an additional
complication compared to the U(1) quantum link model, which is due to the fact
that the Gauss law is realized differently in these two models. This further
complication requires the introduction of so-called Dirac-strings, in order to
consistently relate the HV $h^X_{\widetilde x}$ with the electric fluxes
$E_{x,y}$. These Dirac-strings are located in a staggered fashion on the
vertical links and are denoted by empty squares on the links (see
Fig.~\ref{dimerConfigFluxHeight}).

The quantities $h^{A,B,C,D}$, residing at the sites of a dual sublattice, are
related to the electric flux variables on the links by
\begin{eqnarray} 
E_{x,x+\hat 1} &=& [h^X_{\widetilde x} - h^{X'}_{\widetilde x - \hat 2}] \,
\mbox{mod} 2 = \pm \frac{1}{2},\nonumber \\
E_{x,x+\hat 2} &=& (-1)^{x_1+x_2} [h^X_{\widetilde x} - h^{X'}_{\widetilde x - \hat 1}]
\, \mbox{mod} 2 = \pm \frac{1}{2}, \\ 
&& X,X' \in \{A,B,C,D\}. \nonumber
\end{eqnarray}
Note that whenever $(-1)^{x_1+x_2}= -1$, it indicates the presence 
of a Dirac-string on the relevant vertical link. The corresponding height 
representation and the flux
representation for a columnar dimer configuration is illustrated in
Fig.~\ref{dimerConfigFluxHeight}. Beside the height and flux variables, we
have also marked positive and negative background charges (filled and empty
circles) as well as the Dirac-strings (empty squares).

It should be noted that this construction of height variables is new and fundamentally
different from other height variables definitions that have been introduced in the
literature, in particular, the one described in the review paper of Moessner
and Raman \citep{MR08}. For instance, in our construction the 
height variables take only four values: 0, 1, $\pm 1/2$. On the other hand, 
with the conventional method they are always labelled with integers and can 
take many more values. Our height variables take much fewer values than the
conventional ones, because they provide an exact representation of the dimer
model Hilbert space.

\subsection{Order Parameters and Candidate Phases}
\label{OPandPhases}

In this subsection we review four order parameters in terms of the height
variables that we introduced in \citep{BBHJWW14}. Each of the four candidate
phases --- staggered, columnar, plaquette or mixed --- can then be
unambiguously identified by the specific values these order parameters take in
the different phases.

\begin{figure}
\begin{center}
\includegraphics[scale=1.0]{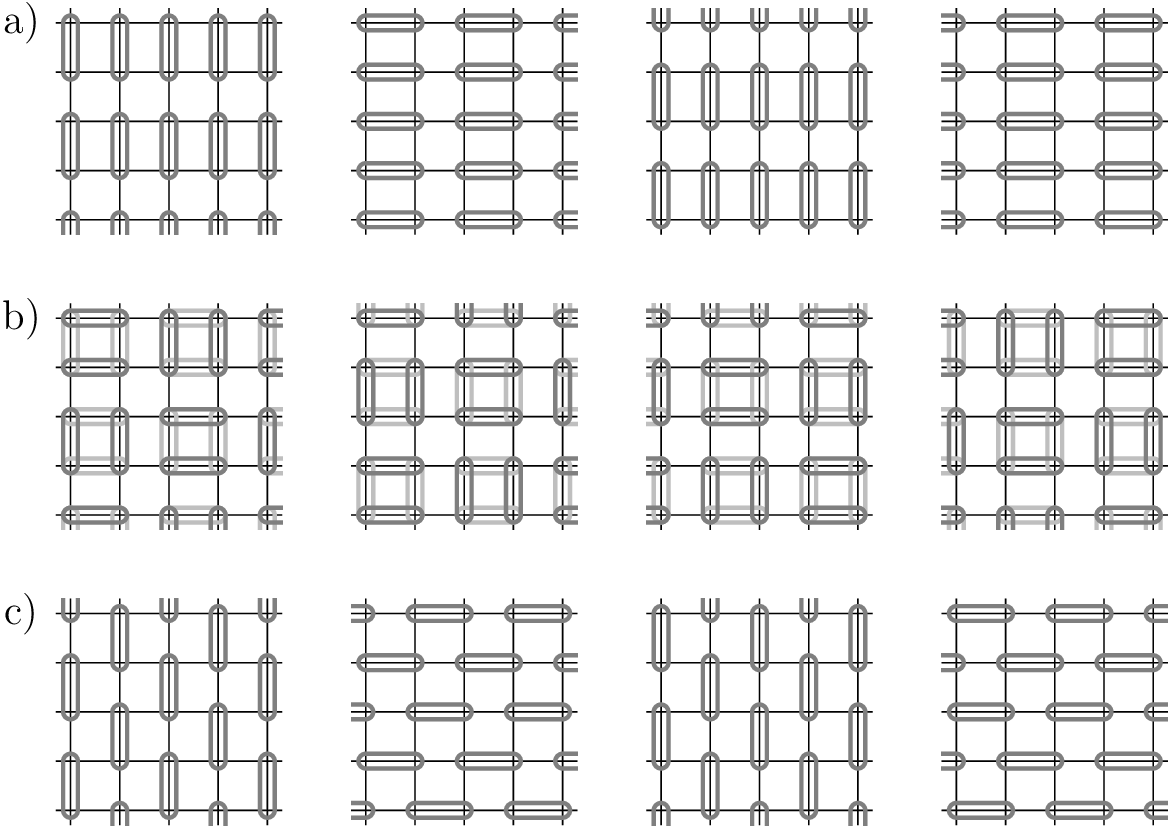} \\ 

\vspace{3mm}

\includegraphics[scale=1.0]{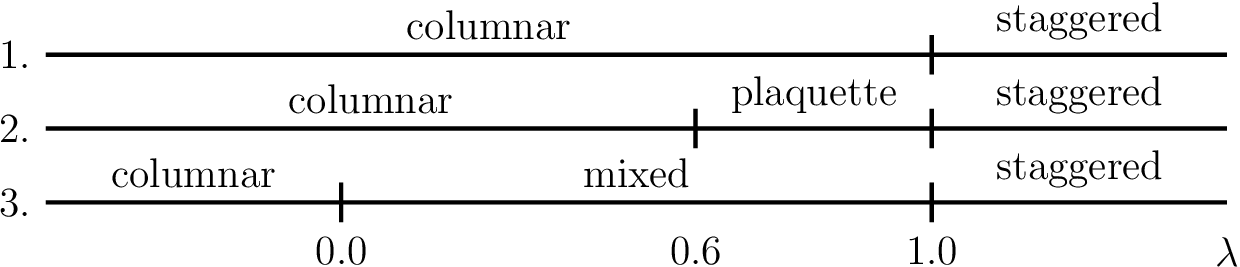}

\end{center}
\caption{Established and conjectured phases for the square
lattice quantum dimer model: (a) Columnar, (b) plaquette, (c) staggered order
on the dual sublattices $A$, $B$, $C$, and $D$. (d) Phase diagram for the
square lattice quantum dimer model as a function of the RK parameter
$\lambda$: three different scenarios \cite{BBHJWW14}.}
\label{FourPhases}
\end{figure}

Let us first discuss the various phases that have been established or
conjectured for the square lattice quantum dimer model. Intuitively, in the
limit $\lambda \rightarrow -\infty$, the system maximizes the number of
flippable plaquettes. On the square lattice, such a state can be obtained by
arranging the dimers in a columnar pattern as depicted in
Fig.~\ref{FourPhases}(a). Note that we are dealing with four-fold degeneracy:
the four columnar configurations are related by translations or rotations.

On the other hand, in the opposite limit $\lambda \rightarrow +\infty$, the
system minimizes the number of flippable plaquettes. On the square lattice,
such a state can be obtained by arranging the dimers in a staggered pattern,
shown in Fig.~\ref{FourPhases}(c). The four degenerate staggered phases are
related by discrete transformations.

Another candidate phase is the so-called plaquette arrangement of dimers,
which is also four-fold degenerate, and is illustrated in
Fig.~\ref{FourPhases}(b). In this phase, pairs of parallel dimers oriented in
both possible directions resonate on the plaquettes belonging to one of the
four sublattices A, B, C, D.

Finally, on the square lattice, another conjectured phase is the so-called
mixed phase which is eight-fold degenerate and corresponds to a superposition
of quantum dimer states, sharing features of both the columnar and the
plaquette phase.

Apart from the established columnar and staggered phases in the limits
$\lambda \rightarrow -\infty$ and $\lambda \rightarrow +\infty$, respectively,
the question of which phases are realized between these two points of
reference --- and what type of associated phase transitions might exist ---
has remained controversial.

In \citep{BBHJWW14} we have challenged the various conflicting scenarios that
have been proposed in earlier studies \citep{Syl06,Ral08} and are depicted in
Fig.~\ref{FourPhases}d.  An important point of reference is the RK point
($\lambda = 1$) where the model is exactly solvable. Away from the RK point, the
situation becomes less clear. Using Green's function Monte Carlo simulations,
the author of Ref.~\citep{Syl06} concludes that on the square lattice there is
a phase transition between the columnar and plaquette phase around
$\lambda \approx 0.6$ (scenario 2 in Fig.~\ref{FourPhases}d). However,
this view is not shared by Ref.~\citep{Ral08} which favors a mixed phase for
$\lambda \gtrsim 0$ according to their Green's function Monte Carlo analysis
(scenario 3 in Fig.~\ref{FourPhases}d). Based on our new order
parameters and a novel Monte Carlo technique we concluded that the system
exists in a columnar phase all the way up to the RK point (scenario 1 in
Fig.~\ref{FourPhases}d). In particular, we found no evidence for plaquette or
mixed phases.

For completeness, we now review the four order parameters whose construction
is based on the dual height representation. Remember that we have two sets of
height variables, the first one associated with the even sublattices A and D,
the second one related to the odd sublattices B and C (see
Fig.~\ref{DualHeightVariables}). 

We first define four auxiliary order parameters $M_A, M_B, M_C, M_D$ as
\begin{equation}
M_X = \sum_{\widetilde x \in X} s^X_{\widetilde x} h^X_{\widetilde x},
\end{equation}
with
\begin{eqnarray}
s^A_{\widetilde x} & = & s^C_{\widetilde x} = (-1)^{({\widetilde x}_1 + \frac{1}{2})/2} ,
\qquad \text{if} \quad {\widetilde x}_1 + \frac{1}{2}
\quad \text{even} ,
\nonumber \\
s^B_{\widetilde x} & = & s^D_{\widetilde x} = (-1)^{({\widetilde x}_1 - \frac{1}{2})/2} ,
\qquad \text{if} \quad {\widetilde x}_1 + \frac{1}{2}
\quad \text{odd} .
\end{eqnarray}
Remember that the height variables on the various sublattices take the values
\begin{equation}
h^{A,D}_{\widetilde x} = 0,1 , \qquad h^{B,C}_{\widetilde x} = \pm \frac{1}{2} .
\end{equation}
We then form the linear combinations,
\begin{eqnarray}
\label{OrderParametersMij}
M_{11} & = & M_A - M_B - M_C + M_D = M_1 \cos\varphi_1, \nonumber \\
M_{22} & = & M_A + M_B - M_C - M_D = M_1 \sin\varphi_1, \nonumber \\
M_{12} & = & M_A - M_B - M_C - M_D = M_2 \cos\varphi_2, \nonumber \\
M_{21} & = & - M_A + M_B - M_C - M_D = M_2 \sin\varphi_2,
\end{eqnarray}
which define the order parameters $M_{11}, M_{12}, M_{21}, M_{22}$ that are more
appropriate to distinguish the phases. The two angles $\varphi_1$ and
$\varphi_2$ define the angle 
\begin{equation}
\varphi = \frac{1}{2}(\varphi_1 + \varphi_2 + \frac{\pi}{4}) .
\end{equation}
In the columnar phase this angle amounts to
$\varphi = 0 \, \mbox{mod} \frac{\pi}{4}$, while in the plaquette phase it
takes the value $\varphi = \frac{\pi}{8} \, \mbox{mod} \frac{\pi}{4}$. Note
that the order parameter values $\pm (M_A,M_B,M_C,M_D)$, and therefore
$\varphi$ and $\varphi + \pi$, represent the same physical dimer
configuration, because a dimer configuration is invariant under a shift of the
height variables,
\begin{equation}
h^X_{\widetilde x}(t)' = [h^X_{\widetilde x}(t) + 1] \, \mbox{mod} 2 .
\end{equation}

\begin{figure}
\begin{center}
\includegraphics[width=12cm]{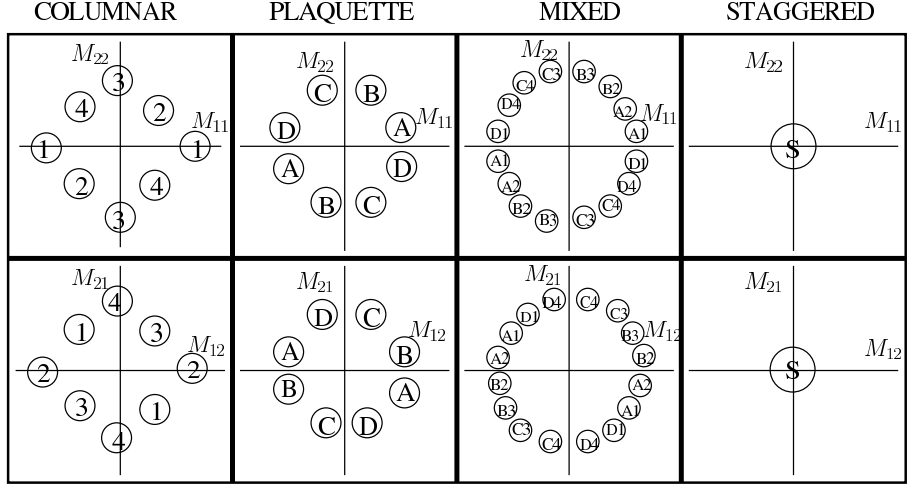}
\end{center}
\caption{The four candidate phases --- columnar, plaquette,
mixed, staggered --- can unambiguously be distinguished by their
characteristic order parameter distributions \cite{BBHJWW14}.}
\label{distributionsOP}
\end{figure}

As illustrated in Fig.~\ref{distributionsOP}, each of the four phases ---
columnar, plaquette, mixed, staggered --- is characterized by its specific
order parameter pattern. While there are four columnar phases (1,2,3,4) and
four plaquette phases (A,B,C,D), there are eight realizations of the mixed
phase (A1,A2,B2,B3,C3,C4,D4,D1). The mixed phases share features of both the
columnar and the plaquette phases. For instance, in a hypothetical phase
transition between a columnar and a mixed phase, a peak in the order parameter
distribution of the columnar phase would split into two individual peaks: the
columnar peak 1 would split into the peaks D1 and A1 referring to the mixed
phase, etc. On the other hand, in a  hypothetical phase transition between a
mixed and a plaquette phase, two peaks in the order parameter distribution of
the mixed phase would merge pairwise into one peak referring to the plaquette
phase: the mixed peaks A1 and A2 would merge into the plaquette peak A, etc.
As we will elaborate in more detail below, in our numerical simulations no
such splitting or merging of peaks is detected. 

For completeness, in Appendix \ref{appendixA} we show how the four order
parameters transform under the symmetries $CT_x, CT_y, O, CO', R_x$, and
$R_y$ of the square lattice quantum dimer model and how the different
symmetries $CT_x, CT_y, O, CO', R_x, R_y$ act on the columnar, plaquette,
and mixed phases, respectively.

It should be pointed out that the specific phases that are actually realized,
depend on both the lattice geometry and the spatial dimension. While we
restrict ourselves to the (two-dimensional) square lattice, more complicated
phases are indeed possible on other lattices. On non-bipartite lattices, in
two or higher spatial dimensions, a $Z_2$ resonance valence bond liquid phase
is formed \citep{MS01,Iva04,RFBIM05}. This is a phase with $Z_2$ topological
order, characterized by four-fold degenerate gapped ground states in the case
of two-dimensional lattices with periodic boundary conditions. In particular,
it has nontrivial excitations and represents a liquid phase because all dimer
correlations decay exponentially. Another example is the U(1) resonating
valence bond liquid phase that is possible on bipartite lattices and in
spatial dimensions three or higher \citep{HKMS03,MS03,Her04}. Even more
complex phases include e.g. the $\sqrt{12} \times \sqrt{12}$ phase, which
appears to be realized on the triangular lattice \citep{MSC00,RFBIM05}.
%Interestingly, while even more complex phases are possible (e.g., on the
%Kagom\'e lattice \citep{ZE95,NS03}), the plaquette phase which is a candidate
%phase of the square lattice, has only been found on two-dimensional bipartite
%lattices \citep{Sac89,Leu96,MSC01,Syl06}.

\subsection{Mean Field Theory}
\label{MeanField}

Using mean field theory, in this subsection we address the question which
phases in the square lattice quantum dimer model may be realized in the
vicinity of the RK point. Following the Ginsburg-Landau-Wilson paradigm, we
formulate an effective action for the system in terms of the order parameters
$M_{11}, M_{12}, M_{21}$, and $M_{22}$, defined in Eq.~(\ref{OrderParametersMij}).
The most general expression up to quartic order that respects all the
symmetries of the underlying quantum dimer model, is given by
\begin{eqnarray}
\label{fullPotential}
&&V = \mu_1 O_1 + \mu_2 O_2 + \nu_0 O_1 O_2 +
\sum_{i=1}^5 \nu_i |O_i|^2, \nonumber \\
&&O_1 = M_{11}^2 + M_{22}^2 + M_{12}^2 + M_{21}^2, \nonumber \\
&&O_2 = M_{11} M_{12} - M_{11} M_{21} + M_{22} M_{12} + M_{22} M_{21}, \nonumber \\
&&O_3 = M_{11}^2 + M_{22}^2 - M_{12}^2 - M_{21}^2, \nonumber \\
&&O_4 = M_{11} M_{12} + M_{11} M_{21} - M_{22} M_{12} + M_{22} M_{21}, \nonumber \\
&&O_5 = M_{11} M_{22} + i M_{12} M_{21}.
\end{eqnarray}
We have two quadratic and six quartic operators, i.e., a total of eight
parameters $\mu_1, \mu_2, \nu_0, \dots, \nu_5$. Each of the terms in the
effective potential is invariant under the discrete symmetries, i.e., under
$O, CO', R_x, R_y, CT_x$ and $CT_y$.

We perform a systematic analysis of the minima of the potential $V$ in the
infinitesimal neighborhood of the staggered phase which begins at the RK
point, and is characterized by $M_{11}$ = $M_{12}$ = $M_{21}$ = $M_{22}$ = 0.
Since the staggered phase corresponds to a stable
minimum, we first diagonalize the mass squared matrix $\mathbf{M}$
\begin{equation}
\mu_1 O_1 + \mu_2 O_2 = (M_{11}, M_{12}, M_{21}, M_{22}) \, \mathbf{M} \,
{(M_{11}, M_{12}, M_{21}, M_{22})}^T
\end{equation}
near this point. All eigenvalues turn out to be positive if the two conditions
\begin{equation}
\mu_1 + \frac{\mu_2}{\sqrt{2}} > 0 , \quad \mu_1 - \frac{\mu_2}{\sqrt{2}} > 0 ,
\end{equation}
are satisfied. Assuming $\mu_1, \mu_2 > 0$, we obtain two zero eigenvalues
if $\mu_2=\sqrt{2}\mu_1$. The corresponding eigenvectors $v_1$ and $v_2$
define the $xy$-plane of vectors $v$ parametrized by
\begin{eqnarray}
\label{flatDirections}
\left(\begin{array}{c}
\frac{x}{\sqrt{2}} \\
\mbox{$\frac{1}{2}$} (y-x) \\
\mbox{$\frac{1}{2}$} (y+x) \\
-\frac{y}{\sqrt{2}} \end{array} \right) .
\end{eqnarray}
This plane corresponds to the flat directions in which the staggered phase is
about to become unstable. Let us therefore evaluate the quartic potential
along these flat directions. The calculation shows that the potential can be
reduced to the simple form
\begin{eqnarray}
V(x,y) & = & (\mu_1 - \frac{\mu_2}{\sqrt{2}}) (x^2 + y^2)
+ \Big( -\frac{\nu_0}{\sqrt{2}}  + \nu_1 + \frac{\nu_2}{2}
+ \frac{\nu_5}{16} \Big) {(x^2 + y^2)}^2 \nonumber \\
& = & \mu (x^2 + y^2) + \nu {(x^2 + y^2)}^2 .
\end{eqnarray}
The minima of the potential $V$ form a circle of radius $r=\sqrt{x^2 + y^2}$
with $r^2=-\mu/2\nu$. Parametrizing the vacuum circle by an angle $\phi$
as $x=r \cos \phi,y= r \sin \phi$, we get
\begin{eqnarray}
\label{minCircle}
\left(\begin{array}{c}
\frac{r}{\sqrt{2}} \cos\phi \\
\frac{r}{2} (\sin\phi - \cos\phi) \\
\frac{r}{2} (\sin\phi + \cos\phi) \\
- \frac{r}{\sqrt{2}} \sin\phi \end{array} \right).
\end{eqnarray}

We now derive general conditions for the minima of the full potential $V$
displayed in Eq.~(\ref{fullPotential}). After a lenghty, but 
otherwise trivial calculation, for the columnar phase, characterized
by $M_{22}=0$ and $M_{21} = -M_{12}$, the potential at a columnar minimum amounts
to
\begin{equation}
\label{minColumnar}
V(M_{11},M_{12},-M_{12},0) = \frac{\mu_1}{2} (M_{11}^2 + 2 M_{12}^2)
+ \mu_2 M_{11} M_{12} .
\end{equation}
An analogous calculation for the plaquette phase, characterized by
$M_{12} = M_{11}$ and $M_{21} = -M_{22}$, shows that the potential at a plaquette
minimum corresponds to
\begin{equation}
\label{minPlaquette}
V(M_{11},M_{11},-M_{22},M_{22}) = \mu_1 (M_{11}^2 + M_{22}^2) + \frac{\mu_2}{2} 
(M_{11}^2 - M_{22}^2 + 2 M_{11} M_{22}) .
\end{equation}
However, it turns out that these two points, along with the six additional
points that correspond to the other columnar and plaquette phases, all have
the same energy on the circle of minima, Eq.~(\ref{minCircle}). Hence, in
order to decide which phase --- columnar or plaquette --- is in fact favored,
we have to perturb around these minima.

A stability analysis shows that there are indeed unstable directions,
associated with negative eigenvalues of the mass squared matrix of the second
derivatives. In fact, both the columnar and the plaquette phase can be
associated with negative eigenvalues and the energy of both phases can be
lowered by proceeding into the unstable directions. However, the relative
energies of the phases reached in this way are very sensitive to the parameters
$\nu_i$ that are unknown. The mean-field analysis hence does not lead to a
conclusive answer of which phase --- columnar or plaquette --- is preferred
near the RK point. Symmetries alone do not favor one of these two candidate
phases over the other. We are thus dealing with a truly dynamical question
which has to be explored with more elaborate methods such as Monte Carlo
simulations (see below).

\section{Low-Energy Spectrum in Finite Volume}
\label{FVES}

In this section, we consider the lowest states in the finite-volume energy
spectrum associated with the columnar, plaquette, and mixed phases,
respectively. This will be useful for identifying the phase structure based on
numerical results obtained by exact diagonalization studies.

\subsection{Low-Energy Spectrum in the Columnar Phase}

Let us first consider the finite-volume energy spectrum in the columnar phase.
The four columnar phases give rise to four almost degenerate eigenstates,
which can be chosen as simultaneous eigenstates of the 90 degrees rotation $O$
with eigenvalues $+1, -i, +i, -1$ as
\begin{eqnarray}
|+1 \rangle & = & \mbox{$\frac{1}{2}$} (|1\rangle +   |2\rangle + |3\rangle
+ |4\rangle) , \nonumber \\
|-i \rangle & = & \mbox{$\frac{1}{2}$} (|1\rangle + i |2\rangle - |3\rangle
- i |4\rangle) , \nonumber \\
|+i \rangle & = & \mbox{$\frac{1}{2}$} (|1\rangle - i |2\rangle - |3\rangle
+ i |4\rangle) , \nonumber \\
|-1 \rangle & = & \mbox{$\frac{1}{2}$} (|1\rangle -   |2\rangle + |3\rangle
- |4\rangle) ,
\end{eqnarray}
with
\begin{eqnarray}
O \, |+1 \rangle & = & |+1 \rangle , \quad O \, |-i \rangle = -i |-i \rangle ,
\nonumber \\
O \, |+i \rangle & = & i |+i \rangle , \quad O \, |-1 \rangle = - |-1 \rangle .
\end{eqnarray}
Under the other discrete symmetries, these states transform as
\begin{eqnarray}
CO' \, |+1 \rangle & = &    |+1 \rangle , \quad CO' \, |-i \rangle = -i |+i
\rangle , \nonumber \\
CO' \, |+i \rangle & = &  i |-i \rangle , \quad CO' \, |-1 \rangle = - |-1
\rangle , \nonumber \\
CT_x \, |+1 \rangle & = &   |+1 \rangle , \quad CT_x \, |-i \rangle =  |+i
\rangle , \nonumber \\
CT_x \, |+i \rangle & = &   |-i \rangle , \quad CT_x \, |-1 \rangle =  |-1
\rangle , \nonumber \\
CT_y \, |+1 \rangle & = &   |+1 \rangle , \quad CT_y \, |-i \rangle = - |+i
\rangle , \nonumber \\
CT_y \, |+i \rangle & = & - |-i \rangle , \quad CT_y \, |-1 \rangle =  |-1
\rangle.
\end{eqnarray}
Besides $|\pm 1 \rangle$, we can also construct linear combinations of
$|\pm i \rangle$ which are eigenstates of $CT_x$ and $CT_y$ such that
\begin{eqnarray}
&&CT_x |+1 \rangle = |+1 \rangle, \nonumber \\
&&CT_x \frac{1}{\sqrt{2}}(|+i \rangle \pm |-i \rangle) = \pm  
\frac{1}{\sqrt{2}}(|+i \rangle \pm |-i \rangle), \nonumber \\
&&CT_x |-1 \rangle = |-1 \rangle, \nonumber \\
&&CT_y |+1 \rangle = |+1 \rangle, \nonumber \\
&&CT_y \frac{1}{\sqrt{2}}(|+i \rangle \pm |-i \rangle) = \mp
\frac{1}{\sqrt{2}}(|+i \rangle \pm |-i \rangle), \nonumber \\
&&CT_y |-1 \rangle = |-1 \rangle.
\end{eqnarray}
This implies that in the columnar phase, in a finite volume there are four
almost degenerate ground states with $(CT_x,CT_y)$ quantum numbers
$(+,+)$, $(+,-)$, $(-,+)$, $(+,+)$. 

\begin{figure}
\begin{center}
\includegraphics[width=12cm]{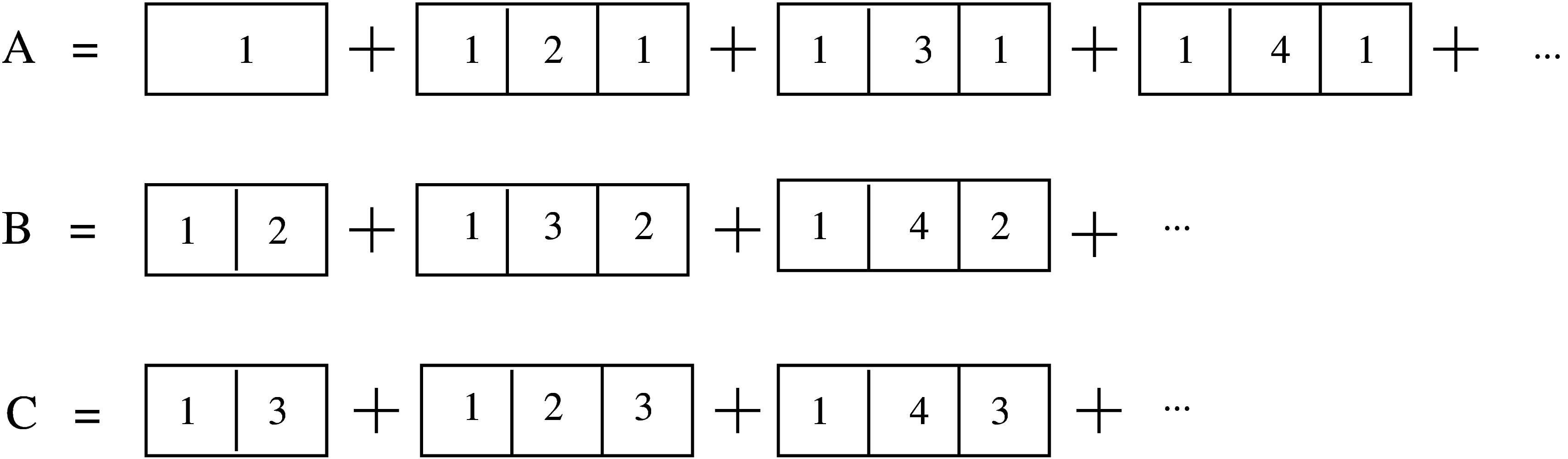}
\end{center}
\caption{A pictorial representation of the transition amplitudes
appearing in Eq.~(\ref{transition}). The numbers 1,~2,~3,~4 represent
the four columnar phases.}
\label{transition1}
\end{figure}

These four states are eigenstates of a reduced transfer matrix
\begin{eqnarray}
\label{transition}
T = \exp(- \beta H)&=&\left(\begin{array}{cccc}
A & B & C & B \\

B & A & B & C \\

C & B & A & B \\

B & C & B & A \end{array} \right) .
\end{eqnarray}
Here, $A, B, C$ are transition amplitudes connecting the various phases. The
corresponding transfer matrix eigenvalues are
\begin{eqnarray}
\exp(-\beta E_{+1}) & = & A + 2B +C , \nonumber \\
\exp(-\beta E_{\pm i}) & = & A - C , \nonumber \\
\exp(-\beta E_{-1}) & = & A - 2B +C .
\end{eqnarray}
Notice only three transition amplitudes appear in 
Eq.~(\ref{transition}). This is because the transitions from $|+1 \rangle$ to 
$|i \rangle$ and $|+1 \rangle$ to $|-i \rangle$ are of the same type. 
A pictorial representation for each transition amplitude is depicted in 
Fig.~\ref{transition1}. Using a dilute instanton gas approximation, 
one can derive analytic
expressions for the transfer matrix elements $A, B, C$. There are instantons
that represent tunneling events between the phases 1 or 3 to 2 or 4. These
instantons have a Boltzmann weight
$\delta_{\perp} \exp(- \alpha_{\perp} L_x L_y)$. In addition, there are instantons
connecting the phases 1 with 3, as well as 2 with 4. These have a Boltzmann
weight $\delta_{\parallel} \exp(- \alpha_{\parallel} L_x L_y)$. The factors
$\delta_{\perp}$ and $\delta_{\parallel}$ describe capillary wave fluctuations of
the instantons. Denoting the free energy in a bulk phase by $f$, an additional
Boltzmann factor $\exp(-\beta f L_x L_y)$ arises as well. The explicit
calculation for $A, B, C$ then yields the following expressions for the
exponentially small energy gaps,
\begin{eqnarray}
E_{\pm i} - E_{+1} & = & 2 \delta_{\perp} \exp(- \alpha_{\perp} L_x L_y)
+ 2 \delta_{\parallel} \exp(- \alpha_{\parallel} L_x L_y) , \nonumber \\
E_{-1} - E_{+1} & = & 4 \delta_{\perp} \exp(- \alpha_{\perp} L_x L_y) .
\end{eqnarray}
In this calculation we have assumed that the interfaces that correspond to a
1-3 (or 2-4) instanton with an interface tension $\alpha_{\parallel}$ are not
completely wet by the other phases 2, 4 (or 1, 3). This assumption implies
$\alpha_{\parallel} < 2 \alpha_{\perp}$. Antonov's rule \citep{Ant07} excludes
$\alpha_{\parallel} > 2 \alpha_{\perp}$, because interfaces with tension
$\alpha_{\parallel}$ would then be unstable against the decay into two
interfaces with tension $\alpha_{\perp}$. This is the situation of complete
wetting. Interfaces with tension $\alpha_{\parallel}$ then simply do not exist
and the corresponding equations turn into
\begin{eqnarray}
E_{\pm i} - E_{+1} & = & 2 \delta_{\perp} \exp(- \alpha_{\perp} L_x L_y) ,
\nonumber \\
E_{-1} - E_{+1} & = & 4 \delta_{\perp} \exp(- \alpha_{\perp} L_x L_y)
= 2 (E_{\pm i} - E_{+1}) .
\end{eqnarray}
Equidistant level spacings are characteristic for complete wetting. At least
for $\lambda \to -\infty$ one indeed expects complete wetting.

\subsection{Low-Energy Spectrum in the Plaquette Phase}

We now consider the finite-volume energy spectrum in the plaquette phase.
Similar to the analysis in the columnar phase, we define the four plaquette
eigenstates as
\begin{eqnarray}
|+1 \rangle' & = & \mbox{$\frac{1}{2}$} (|A\rangle +   |B\rangle + |C\rangle
+   |D\rangle) , \nonumber \\
|-i \rangle' & = & \mbox{$\frac{1}{2}$} (|A\rangle + i |B\rangle - |C\rangle
- i |D\rangle) , \nonumber \\
|+i \rangle' & = & \mbox{$\frac{1}{2}$} (|A\rangle - i |B\rangle - |C\rangle
+ i |D\rangle) , \nonumber \\
|-1 \rangle' & = & \mbox{$\frac{1}{2}$} (|A\rangle -   |B\rangle + |C\rangle
-   |D\rangle) .
\end{eqnarray}
Under the discrete symmetries they transform as
\begin{eqnarray}
O \, |+1 \rangle' & = &  |+1 \rangle' , \quad O \, |-i \rangle'
= -i |-i \rangle' , \nonumber \\
O \, |+i \rangle' & = &  i |+i \rangle' , \quad O \, |-1 \rangle'
= - |-1 \rangle', \nonumber \\
CO' \, |+1 \rangle' & = &    |+1 \rangle' , \quad CO' \, |-i \rangle'
= - |+i \rangle' , \nonumber \\
CO' \, |+i \rangle' & = &  - |-i \rangle' , \quad CO' \, |-1 \rangle'
= |-1 \rangle' , \nonumber \\
CT_x \, |+1 \rangle' & = &   |+1 \rangle' , \quad CT_x \, |-i \rangle'
=  -i |+i \rangle' , \nonumber \\
CT_x \, |+i \rangle' & = &   i |-i \rangle' , \quad CT_x \, |-1 \rangle'
 =  - |-1 \rangle' , \nonumber \\
CT_y \, |+1 \rangle' & = &   |+1 \rangle' , \quad CT_y \, |-i \rangle'
= i |+i \rangle' , \nonumber \\
CT_y \, |+i \rangle' & = & -i |-i \rangle' , \quad CT_y \, |-1 \rangle'
= - |-1 \rangle'.
\end{eqnarray}
Besides $|\pm 1 \rangle'$, we can also construct linear combinations of
$|\pm i \rangle'$ which are eigenstates of $CT_x$ and $CT_y$ such that
\begin{eqnarray}
&&CT_x |+1 \rangle' = |+1 \rangle', \nonumber \\
&&CT_x \frac{1}{\sqrt{2}}(|+i \rangle' \pm i |-i \rangle') = \pm  
\frac{1}{\sqrt{2}}(|+i \rangle' \pm i |-i \rangle'), \nonumber \\
&&CT_x |-1 \rangle' = - |-1 \rangle', \nonumber \\
&&CT_y |+1 \rangle' = |+1 \rangle', \nonumber \\
&&CT_y \frac{1}{\sqrt{2}}(|+i \rangle' \pm i |-i \rangle') = \mp  
\frac{1}{\sqrt{2}}(|+i \rangle' \pm i |-i \rangle'), \nonumber \\
&&CT_y |-1 \rangle' = - |-1 \rangle'.
\end{eqnarray}
Like in the columnar phase, in the plaquette phase there are four almost 
degenerate ground states. However, in contrast to the columnar phase, their
$(CT_x,CT_y)$ quantum numbers are $(+,+)$, $(+,-)$, $(-,+)$, $(-,-)$. In
particular, the quantum numbers of the third excited state are different in
the two cases.

The calculation of the energy spectrum in the plaquette phase is the same as
in the columnar phase and shall not be repeated here.

\subsection{Low-Energy Spectrum in the Mixed Phase}

Finally we discuss the lowest states in the finite-volume energy spectrum of
the mixed phase. Here we have a total of eight states that become degenerate
in the infinite volume limit, with exponentially small gaps at finite volume.
Let us construct the states as eigenstates of an explicitly broken and thus
only approximate $\Z(8)$ symmetry. Defining $z = \exp(2 \pi i/8)$, we obtain
\begin{eqnarray}
|+1 \rangle'' & = & \frac{1}{\sqrt{8}} \Big( |A_1\rangle + |A_2\rangle
+ |B_2\rangle + |B_3\rangle
+ |C_3\rangle + |C_4\rangle + |D_4\rangle + |D_1\rangle \Big) , \nonumber \\
|z \rangle'' & = & \frac{1}{\sqrt{8}} \Big( |A_1\rangle + z |A_2\rangle
+ i |B_2\rangle + z^3 |B_3\rangle
- |C_3\rangle + z^5 |C_4\rangle - i |D_4\rangle + z^7 |D_1\rangle \Big) ,
\nonumber \\
|+i \rangle'' & = & \frac{1}{\sqrt{8}} \Big( |A_1\rangle + i |A_2\rangle
- |B_2\rangle -i |B_3\rangle
+ |C_3\rangle + i |C_4\rangle - |D_4\rangle -i |D_1\rangle \Big) , \nonumber \\
|z^3 \rangle'' & = & \frac{1}{\sqrt{8}} \Big( |A_1\rangle + z^3 |A_2\rangle
- i |B_2\rangle + z |B_3\rangle
- |C_3\rangle + z^7 |C_4\rangle + i |D_4\rangle + z^5 |D_1\rangle \Big) ,
\nonumber \\
|-1 \rangle'' & = & \frac{1}{\sqrt{8}} \Big( |A_1\rangle - |A_2\rangle
+ |B_2\rangle - |B_3\rangle
+ |C_3\rangle - |C_4\rangle + |D_4\rangle - |D_1\rangle \Big) , \nonumber \\
|z^5 \rangle'' & = & \frac{1}{\sqrt{8}} \Big( |A_1\rangle + z^5 |A_2\rangle
+ i |B_2\rangle + z^7 |B_3\rangle
- |C_3\rangle + z |C_4\rangle - i |D_4\rangle + z^3 |D_1\rangle \Big) ,
\nonumber \\
|-i \rangle'' & = & \frac{1}{\sqrt{8}} \Big( |A_1\rangle -i |A_2\rangle
- |B_2\rangle +i |B_3\rangle
+ |C_3\rangle -i |C_4\rangle - |D_4\rangle + i |D_1\rangle \Big) , \nonumber \\
|z^7 \rangle'' & = & \frac{1}{\sqrt{8}} \Big( |A_1\rangle + z^7 |A_2\rangle
- i |B_2\rangle + z^5 |B_3\rangle
- |C_3\rangle + z^3 |C_4\rangle + i |D_4\rangle + z |D_1\rangle \Big) .\nonumber \\
\end{eqnarray}
This gives rise to the following transformation rules,
\begin{eqnarray}
O \, |+1 \rangle'' & = &   |+1 \rangle'' , \quad O \, |z \rangle''   
= -i |z \rangle'' , \nonumber \\
O \, |+i \rangle'' & = & - |+i \rangle'' , \quad O \, |z^3 \rangle'' 
=  i |z^3 \rangle'', \nonumber \\
O \, |-1 \rangle'' & = &   |-1 \rangle'' , \quad O \, |z^5 \rangle''
= -i |z^5 \rangle'' , \nonumber \\
O \, |-i \rangle'' & = & - |-i \rangle'' , \quad O \, |z^7 \rangle'' 
=  i |z^7 \rangle'', \nonumber \\
CO' \, |+1 \rangle'' & = &    |+1 \rangle'' , \quad CO' \, |z \rangle''   
= z^5 |z^7 \rangle'' , \nonumber \\
CO' \, |+i \rangle'' & = &  i |-i \rangle'' , \quad CO' \, |z^3 \rangle'' 
= z^7 |z^5 \rangle'', \nonumber \\
CO' \, |-1 \rangle'' & = &  - |-1 \rangle'' , \quad CO' \, |z^5 \rangle'' 
= z   |z^3 \rangle'' , \nonumber \\
CO' \, |-i \rangle'' & = & -i |+i \rangle'' , \quad CO' \, |z^7 \rangle'' 
= z^3 |z \rangle'', \nonumber \\
CT_x \, |+1 \rangle'' & = &     |+1 \rangle'' , \quad CT_x \, |z \rangle''   
= z^7 |z^7 \rangle'' , \nonumber \\
CT_x \, |+i \rangle'' & = &  -i |-i \rangle'' , \quad CT_x \, |z^3 \rangle'' 
= z^5 |z^5 \rangle'', \nonumber \\
CT_x \, |-1 \rangle'' & = &   - |-1 \rangle'' , \quad CT_x \, |z^5 \rangle''
= z^3 |z^3 \rangle'' , \nonumber \\
CT_x \, |-i \rangle'' & = &   i |+i \rangle'' , \quad CT_x \, |z^7 \rangle'' 
= z   |z \rangle'', \nonumber \\
CT_y \,  |+1 \rangle'' & = &     |+1 \rangle'' , \quad CT_y \, |z \rangle''   
= z^3 |z^7 \rangle'' , \nonumber \\
CT_y \,  |+i \rangle'' & = &  -i |-i \rangle'' , \quad CT_y \, |z^3 \rangle'' 
= z  |z^5 \rangle'', \nonumber \\
CT_y \,  |-1 \rangle'' & = &   - |-1 \rangle'' , \quad CT_y \, |z^5 \rangle'' 
= z^7 |z^3 \rangle'' , \nonumber \\
CT_y \,  |-i \rangle'' & = &   i |+i \rangle'' , \quad CT_y \, |z^7 \rangle'' 
= z^5 |z \rangle'' .
\end{eqnarray}
By construction, the eight states are eigenstates of the approximate continuous
$U(1)$ symmetry $U$ restricted to $\Z(8)$,
\begin{eqnarray}
U \, |+1 \rangle'' & = &    |+1 \rangle'' , \quad U \, |z \rangle''   
=  z |z \rangle'' , \nonumber \\
U \, |+i \rangle'' & = &  i |+i \rangle'' , \quad U \, |z^3 \rangle'' 
= z^3 |z^3 \rangle'', \nonumber \\
U \, |-1 \rangle'' & = &  - |-1 \rangle'' , \quad U \, |z^5 \rangle'' 
= z^5 |z^5 \rangle'' , \nonumber \\
U \, |-i \rangle'' & = & -i |-i \rangle'' , \quad U \, |z^7 \rangle'' 
= z^7 |z^7 \rangle''.
\end{eqnarray}

The symmetries $CT_x, CT_y, CO'$ have $|+1 \rangle''$ and $|-1 \rangle''$
unmixed, and they mix $|+i \rangle''$ with $|-i \rangle''$, $|z \rangle''$
with $|z^7 \rangle'' = |z^* \rangle''$, and $|z^3 \rangle''$ with
$|z^5 \rangle'' = |z^{3*} \rangle''$. The energy spectrum will thus contain two
non-degenerate states, as well as three pairs of two-fold degenerate states.
We do not explicitly work out the energy spectrum, but point out that in a
mixed phase eight finite-volume states become degenerate in the infinite
volume limit, while for the columnar or plaquette phase only four states
become degenerate.

\section{Low-Energy Effective Theory}
\label{LEEFT}

In \citep{BBHJWW14} we found strong numerical evidence for an emergent soft 
pseudo-Goldstone mode at and below the RK point. In this chapter we discuss
the theoretical concepts underlying our numerical analysis that we present in
the next section. These include the effective field description of the
pseudo-Goldstone boson mode and the energy spectrum in a finite volume.

\subsection{Goldstone Boson Fields, Symmetries, Lagrangian}

The basic degree of freedom in the effective theory --- the soft
pseudo-Goldstone mode --- is parametrized by the angle 
$\varphi = \frac{1}{2}(\varphi_1 + \varphi_2 + \frac{\pi}{4})$.
Note that the angles $\varphi_1$ and $\varphi_2$ have been defined in
Eq.~(\ref{OrderParametersMij}). Under the various symmetries, the angle
$\varphi$ transforms as
\begin{eqnarray}
&&^{CT_x}\varphi = \pi - \varphi, \quad
^{CT_y}\varphi = \frac{\pi}{2} - \varphi, \nonumber \\
&&^O\varphi = \frac{\pi}{4} + \varphi, \quad
^{CO'}\varphi = - \frac{\pi}{4} - \varphi.
\end{eqnarray}
The leading Euclidean effective Lagrangian takes the form
\begin{equation}
{\cal L} = \frac{\rho}{2} \Big( \frac{1}{c^2} \partial_t \varphi \partial_t
\varphi + \partial_i \varphi \partial_i \varphi \Big)
+ \kappa (\partial_i \partial_i \varphi)^2 + \delta \cos^2(4 \varphi) ,
\end{equation}
which is identical with the effective Lagrangian of the $(2+1)$-dimensional
$\RP(1)$ model. Note that the angles $\varphi$ and $\varphi + \pi$ are
indistinguishable, such that the physical Hilbert space only contains states
that are invariant under this shift.

While $\rho$ is the spin stiffness, the quantity $c$ is the limiting velocity
of the pseudo-Goldstone boson. The term proportional to the low-energy
effective constant $\delta$ explicitly breaks the emergent $SO(2)$ symmetry to
the discrete subgroup $\Z(8)$. Accordingly, we are dealing with a light
pseudo-Goldstone mode with mass
\begin{equation}
M c = 4 \sqrt{2 |\delta|/ \rho } .
\end{equation}

At the RK point ($\lambda=1$) all flux configurations cost zero energy in
their ground state. This implies that the individual effective couplings
$\delta$ and $\rho$ (but not the ratio $\rho/c^2)$ are zero. Note that the
condition $\rho=0$ at the RK point is an analytic result which does not
require any fine-tuning. In this case, the quartic kinetic term proportional
to $\kappa$ becomes the dominant contribution. Remarkably,
$\partial_i \partial_i \varphi = 0$ for all configurations of static external
charges in their ground state, such that this term indeed does not contribute
any ground state energy. This is not true for the term
$\sum_{i=1,2}\partial_i \partial_i \varphi\partial_i \partial_i \varphi$. Hence,
such a term cannot arise at the RK point, but it can arise away from it. Also
all terms in the potential energy vanish at the RK point.

\subsection{Rotor Spectrum}

Let us first consider the spectrum of vacuum states in a periodic volume
$L_1 \times L_2$ at $\delta=0$. At zero temperature, to lowest order, we may
assume $\varphi(x,t) = \varphi(t)$, i.e. the low-energy dynamics reduces to the
one of the spatial zero-mode, which represents a single quantum mechanical 
degree of freedom. The action then reduces to
\begin{equation}
S[\varphi] = \int dt \, \Bigg[ \frac{\rho L_1 L_2}{2 c^2} \, \partial_t
\varphi \partial_t \varphi + \delta L_1 L_2 \cos^2(4 \varphi) \Bigg] \, ,
\end{equation}
and the corresponding quantum mechanical Hamilton operator is given by
\begin{equation}
H_{\rm eff} = -\frac{c^2}{2 \rho L_1 L_2} \, \partial^2_{\varphi} + \delta L_1 L_2
\cos^2(4 \varphi) \, .
\end{equation}
At $\delta=0$, this describes a free ``particle'' on a circle. The
corresponding energy eigenstates and eigenvalues are
\begin{equation}
\psi_m(\varphi) = \frac{1}{\sqrt{2 \pi}} \exp(i m \varphi) \, ,
\qquad E_m = \frac{m^2 c^2}{2 \rho L_1 L_2}.
\end{equation}
Since $\varphi$ and $\varphi + \pi$ are physically equivalent, $m$ is
restricted to even integers.

Let us consider the effects of small $\delta$ in perturbation theory. The
ground state with $m=0$ is non-degenerate and has a constant wavefunction,
\begin{equation}
\psi_0(\varphi) = \frac{1}{\sqrt{2 \pi}} \, .
\end{equation}
Its energy shift is
\begin{equation}
E^{(1)}_0 = \delta L_1 L_2 \langle \psi_0 | \cos^2(4 \varphi) | \psi_0 \rangle 
= \frac{\delta L_1 L_2}{2} \, .
\end{equation}
The excited states with $m = \pm 2, \pm 4, \dots $ are 2-fold degenerate.
Their energy shifts result from
\begin{eqnarray}
V_{m,m} & = & V_{-m,-m} = \frac{\delta L_1 L_2}{2} \, , \\
V_{m,-m} & = & V_{-m,m} = \frac{\delta L_1 L_2}{16 \pi} 
{\Bigg[ \frac{\sin(2(m-4)\varphi)}{m-4} + \frac{2 \sin(2 m \varphi)}{m} 
+ \frac{\sin(2(m+4)\varphi)}{m+4} \Bigg]}_0^{2\pi} \, . \nonumber 
\end{eqnarray}
The case $m = \pm 4$ thus needs to be considered separately. Since
\begin{equation}
\lim_{m\to\pm4} V_{m,-m} = \frac{\delta L_1 L_2}{4} \, ,
\end{equation}
the corresponding energy shift takes the form
\begin{equation}
E^{(1)}_{4 \pm} = \frac{\delta L_1 L_2}{2} \pm \frac{\delta L_1 L_2}{4} \, ,
\end{equation}
and the previously degenerate energy levels split,
\begin{equation}
E_{\pm 4} - E_0 = \frac{8 c^2}{\rho L_1 L_2} \pm
\frac{\delta L_1 L_2}{4} \, .
\end{equation}
This formula is only valid in the regime
\begin{equation}
\delta L_1 L_2 \ll \frac{c^2}{\rho L_1 L_2} \, .
\end{equation}
For $m = \pm 2, \pm 6, \pm 8, \dots$, there is no such effect and we simply
have
\begin{equation}
E_{\pm m} - E_0 = \frac{m^2 c^2}{2\rho L_1 L_2} \, .
\end{equation}

We now turn to second order perturbation theory in $\delta$. Note that the
leading order correction to the higher excited states $m=\pm 2, \pm 4, \dots$
only arises at order $\delta^2$. To avoid degenerate perturbation theory, we
separately consider even and odd wave functions. We begin with the even wave
functions
\begin{equation}
\psi^e_m(\varphi) = \frac{1}{\sqrt{\pi}} \cos(m \varphi) \, , \qquad E_m
 = \frac{m^2 c^2}{2 \rho L_1 L_2} \, , \qquad
\psi^e_0(\varphi) = \frac{1}{\sqrt{2\pi}} \, .
\end{equation}
With the matrix elements, $m,n > 0$,
\begin{equation}
\langle m | V(\varphi) | n \rangle = \frac{\delta L_1 L_2}{\pi} \, 
\int_0^{2 \pi} d \varphi \cos(m \varphi) \cos^2(4 \varphi)
\cos(n \varphi) \, ,
\end{equation}
we obtain
\begin{eqnarray}
E^{even,(2)}_0 & = & - \sum_{n\neq0}  \frac{{| \langle 0 | V(\varphi) | n 
\rangle |}^2}{E_n -E_0}
= - \frac{{| \langle 0 | V(\varphi) | 8 \rangle |}^2}{E_8 -E_0}
= - \frac{\delta^2 \rho L^3_1 L^3_2}{256 c^2}
\, , \nonumber \\
E^{even,(2)}_2 & = & - \sum_{n\neq2}  \frac{{| \langle 2 | V(\varphi) | n 
\rangle |}^2}{E_n -E_2} 
= - \frac{{| \langle 2 | V(\varphi) | 6 \rangle |}^2}{E_6 -E_2}\
- \frac{{| \langle 2 | V(\varphi) | 10 \rangle |}^2}{E_{10} -E_2}
= - \frac{\delta^2 \rho L^3_1 L^3_2}{192 c^2}
 \, , \nonumber \\
E^{even,(2)}_4 & = & - \sum_{n\neq4}  \frac{{| \langle 4 | V(\varphi) | n 
\rangle |}^2}{E_n -E_4} 
= - \frac{{| \langle 4 | V(\varphi) | 12 \rangle |}^2}{E_{12} -E_4}
= - \frac{\delta^2 \rho L^3_1 L^3_2}{1024 c^2}
\, .
\end{eqnarray}
Analogously, for the odd wave functions ($m,n > 0$),
\begin{equation}
\psi^o_m(\varphi) = \frac{1}{\sqrt{\pi}} \sin(m \varphi) \, , 
\qquad E_m = \frac{m^2 c^2}{2 \rho L_1 L_2} \, ,
\end{equation}
and with
\begin{equation}
\langle m | V(\varphi) | n \rangle = \frac{\delta L_1 L_2}{\pi} \, 
\int_0^{2 \pi} d \varphi \sin(m \varphi) \cos^2(4 \varphi)
\sin(n \varphi) \, ,
\end{equation}
second order perturbation theory leads to
\begin{eqnarray}
E^{odd,(2)}_2 & = & - \sum_{n\neq2}  \frac{{| \langle 2 | V(\varphi) | n 
\rangle |}^2}{E_n -E_2}
= - \frac{{| \langle 2 | V(\varphi) | 6 \rangle |}^2}{E_6 -E_2}
- \frac{{| \langle 2 | V(\varphi) | 10 \rangle |}^2}{E_{10} -E_2}
= - \frac{\delta^2 \rho L^3_1 L^3_2}{192 c^2}
 \, , \nonumber \\
E^{odd,(2)}_4 & = & - \sum_{n\neq4}  \frac{{| \langle 4 | V(\varphi) | n 
\rangle |}^2}{E_n -E_4}
= - \frac{{| \langle 4 | V(\varphi) | 12 \rangle |}^2}{E_{12} -E_4}
= - \frac{\delta^2 \rho L^3_1 L^3_2}{1024 c^2}
\, .
\end{eqnarray}

We now proceed with a nonperturbative treatment of $\delta$ and consider the
nonperturbative Schr\"odinger equation that takes the form of a Hill equation,
\begin{eqnarray}
& & -\frac{1}{2} \delta^2_{\varphi} \psi(\varphi) + V_0 \cos^2(4 \varphi)
\psi(\varphi) = \varepsilon \psi(\varphi) \, , \nonumber \\
& & V_0=\frac{\delta \rho L^2_1 L^2_2}{c^2} \, , \qquad \varepsilon
= \frac{\rho L_1 L_2}{c^2} \, E \, .
\end{eqnarray}
Since $\varphi$ and $\varphi + \pi$ are to be identified, we introduce a new
angle as $\varphi' = 4 \varphi$, with $\varphi'$ having period $4 \pi$. The
above equation is then converted into the following Mathieu equation,
\begin{equation}
- \delta^2_{\varphi'} \psi(\varphi') + \frac{V_0}{16} \, \cos(2 \varphi')
\psi(\varphi') = \Bigg( \frac{\varepsilon}{8} -\frac{V_0}{16}
\Bigg) \psi(\varphi') \, .
\end{equation}
The corresponding solutions are even and odd Mathieu functions
\begin{equation}
\psi_{2m}(\varphi') = \frac{1}{\sqrt{\pi}} \, ce_m (\varphi') \, , 
\quad \psi_{2m+1}(\varphi') = \frac{1}{\sqrt{\pi}} \, se_m (\varphi')
\, ,
\end{equation}
with eigenvalue $\lambda_0$ given to lowest order by
\begin{equation}
\lambda_0 = \frac{\varepsilon_0}{8} - \frac{V_0}{16} 
= -\frac{1}{2} {\Bigg( \frac{V_0}{32} \Bigg)}^2 + {\cal O}(V^4_0) \, .
\end{equation}
Accordingly, the ground state energy reads
\begin{equation}
E_0 = \frac{\delta L_1 L_2}{2} - \frac{\delta^2 \rho L^3_1 L^3_2}{256 c^2} 
+ {\cal O}(\delta^4) \, ,
\end{equation}
in agreement with the leading perturbative results. In the nonperturbative
regime, for the excited states we obtain 
\begin{equation}
E_m = \frac{c^2}{\rho L_1 L_2} \varepsilon_m =  \frac{\delta L_1 L_2}{2} 
+ \frac{8c^2}{\rho L_1 L_2} \, \lambda_m\Big(
\frac{\delta \rho L^2_1 L^2_2}{16c^2} \Big) \, .
\end{equation}
The energy splittings in the rotor spectrum are thus given by the eigenvalues
\begin{equation}
\lambda_m = \lambda_m\Big(\frac{\delta \rho L^2_1 L^2_2}{16c^2}\Big)
\end{equation}
of the Mathieu equation,
\begin{equation}
E_m - E_0 = \frac{8 c^2}{\rho L_1 L_2} \, \Bigg[ \lambda_m\Big( \frac{\delta 
\rho L_1^2 L_2^2}{16c^2} \Big)
 - \lambda_0\Big( \frac{\delta \rho L_1^2 L_2^2}{16c^2}\Big) \Bigg] \, .
\end{equation}
As a consistency check we also consider the eigenvalue $\lambda_1$ that
corresponds to the odd Mathieu function $se_1(\varphi')$,
\begin{equation}
\lambda_1 = \frac{\varepsilon_1}{8} - \frac{V_0}{16} = 1 - \frac{V_0}{32} 
- \frac{1}{8} {\Bigg( \frac{V_0}{32} \Bigg)}^2
+ \frac{1}{64} {\Bigg( \frac{V_0}{32} \Bigg)}^3 + {\cal O}(V^4_0) \, .
\end{equation}
Accordingly, the energy $E_1$ is given by
\begin{equation}
E_1 = \frac{8c^2}{\rho L_1 L_2} + \frac{\delta L_1 L_2}{4} 
- \frac{\delta^2 \rho L^3_1 L^3_2}{1024 c^2} 
+ \frac{\delta^3 \rho^2 L^5_1 L^5_2}{262144 c^4} + {\cal O}(\delta^4) \, ,
\end{equation}
such that
\begin{equation}
E_1 - E_0 = \frac{8c^2}{\rho L_1 L_2} - \frac{\delta L_1 L_2}{4} 
+ \frac{3 \delta^2 \rho L^3_1 L^3_2}{1024 c^2} 
+ \frac{\delta^3 \rho^2 L^5_1 L^5_2}{262144 c^4} + {\cal O}(\delta^4) \, .
\end{equation}
This is consistent with the leading order perturbative calculation, when one
identifies the state corresponding to $se_1(\varphi')$ with the state $m=-4$.
Note that the results of second order perturbation theory in $\delta$ are also
consistent with the expansion of the Mathieu function eigenvalues.

The theoretical results derived in this section will be compared with exact
diagonalization and Monte Carlo simulation results in sections
\ref{ExactDiag} and \ref{MC}, respectively.

\section{Exact Diagonalization Results}
\label{ExactDiag}

In this section we discuss exact diagonalization results for $L_1 \times L_2$ 
lattices with $L_1,L_2 \in \{ 4,6,8 \}$, which allow us to determine some 
low-energy parameters of the effective field theory discussed in the previous
section.

\begin{figure}[htbp]
\begin{center}
\includegraphics[scale=1.0]{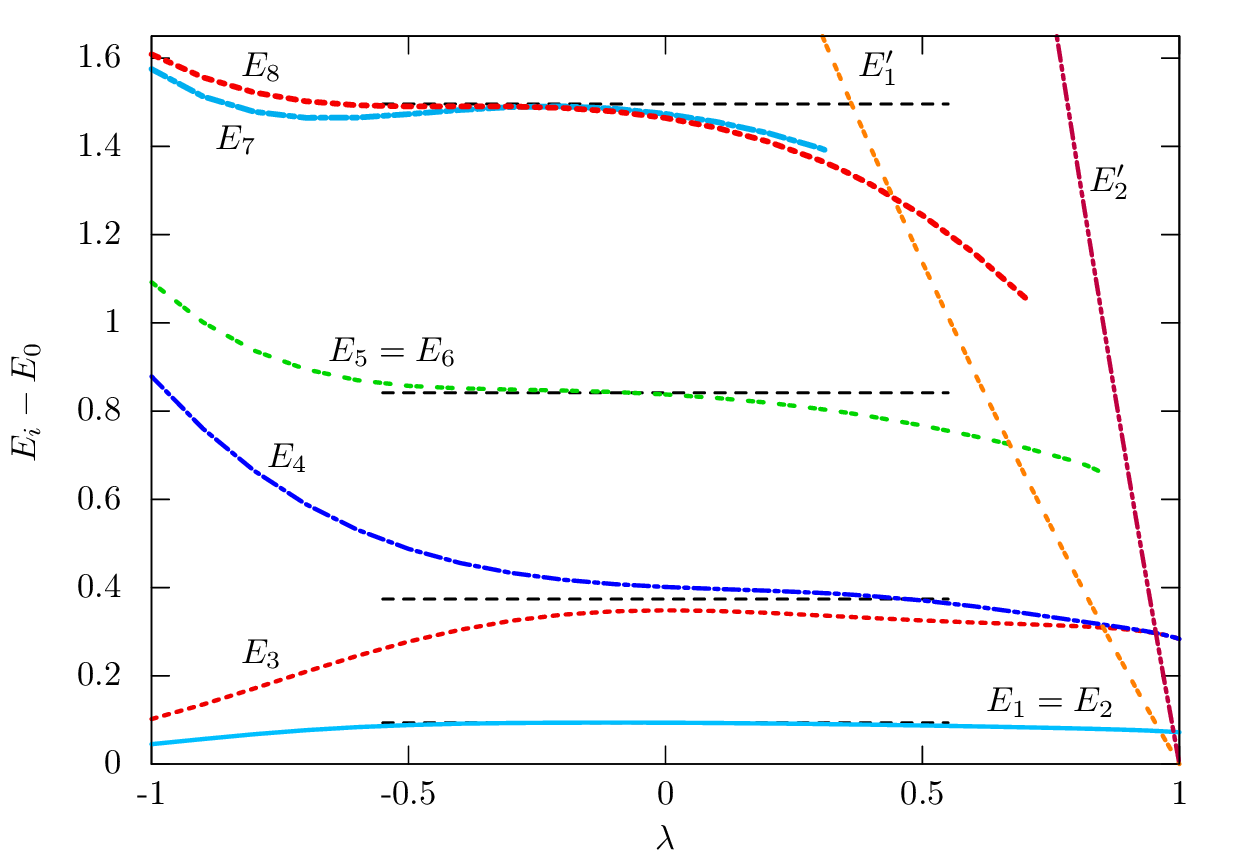}
\end{center}
\caption{[Color online] Energy spectrum on an $8^2$ lattice as a 
function of the RK coupling $\lambda$ \cite{BBHJWW14}.}
\label{statecrossing}
\end{figure}
\begin{figure}[htbp]
\begin{center}
\includegraphics[scale=1.0]{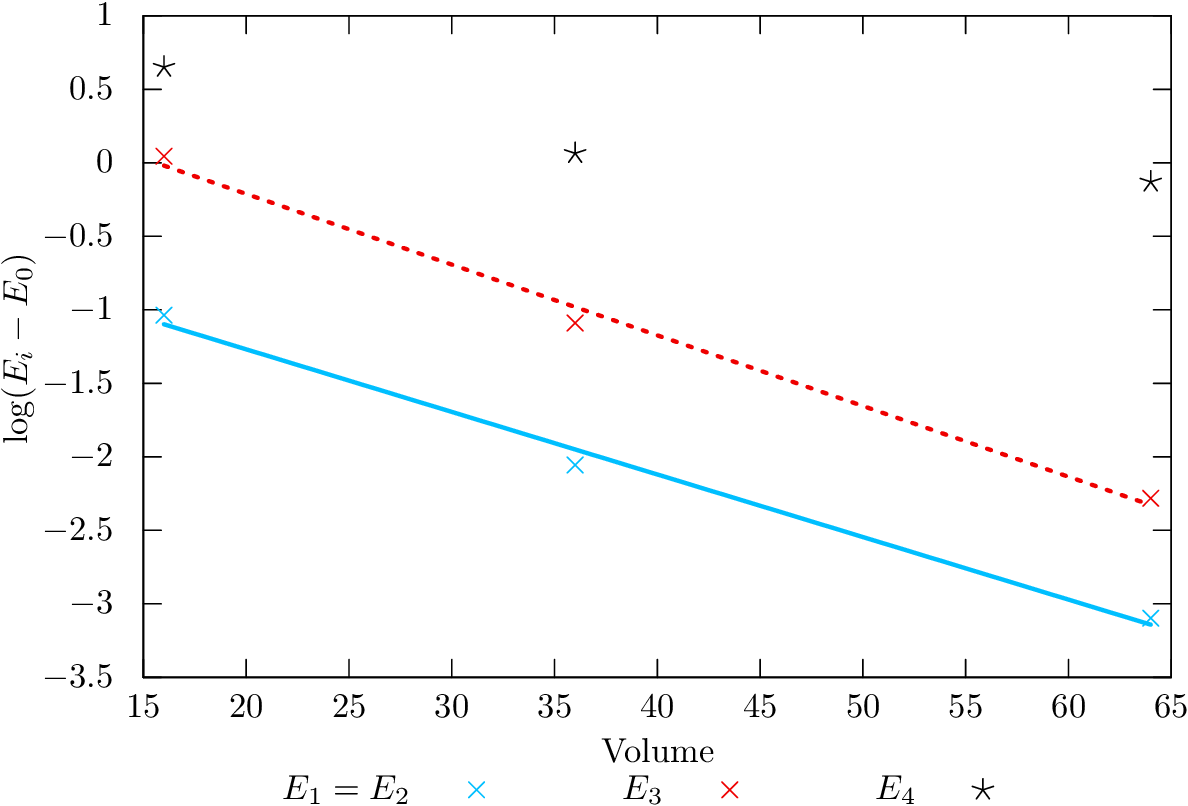}
\end{center}
\caption{[Color online] Logarithmic energy gaps as a function of the
volume at $\lambda=-1$.}
\label{energydiff}
\end{figure}
Using exact diagonalization we were able to calculate the low-lying energy
spectrum on lattices up to $8 \times 8$. Fig.~\ref{statecrossing} shows the
energy gaps on the largest lattice. For $\lambda < 1$, the ground state is
non-degenerate and transforms trivially under the symmetry operations, i.e.\
it has quantum numbers ($CT_x,CT_y$) = (+,+). The first two excited states
with energy gap $E_1 = E_2$ are degenerate and have quantum numbers $(+,-)$
and $(-,+)$, while the next excited state with energy gap $E_3$ has again
quantum numbers $(+,+)$. As Fig.~\ref{energydiff} shows for $\lambda=-1$, the
energy gaps of these three excited states decrease exponentially with the
volume $L_1 L_2$, i.e.\ $E_{1,2}, E_3 \sim \exp(- \alpha L_1 L_2)$ for
$-0.2 \lesssim \lambda \lesssim 0.8$. The fact that the gap between the
finite-volume ground state and the three first excited states is exponentially
small indicates that four phases coexist at zero temperature. The
$(CT_x,CT_y)$ quantum numbers $(+,+)$, $(+,-)$, $(-,+)$, $(+,+)$ indicate that
we are in a columnar and not in a plaquette phase.

If the columnar phase were replaced by the plaquette phase for larger values
of $\lambda$, one would expect a level crossing of the excited $(+,+)$ state
with the lowest $(-,-)$ state. Interestingly, no such level crossing arises in
our exact diagonalization study. Notably, the next excited state with energy
gap $E_4$, does not decrease exponentially with the volume. It has quantum
numbers $(-,-)$ and almost degenerates with the $(+,+)$ state with energy
$E_3$ for $-0.2 \lesssim \lambda \lesssim 0.8$. Furthermore, the next two
states with energy $E_5 = E_6$ are exactly degenerate and again have quantum
numbers $(+,-)$ and $(-,+)$. The next states, with energies $E_7$ and $E_8$
are once more almost degenerate and transform as $(+,+)$ and $(-,-)$. The
energy ratios of these states are  given by
$E_{1,2}:E_{3,4}:E_{5,6}:E_{7,8} \approx 1:4:9:16$, which is indicated by the
dashed lines  in Fig.~\ref{statecrossing}. This hints at an approximate rotor
spectrum. Indeed in \citep{BBHJWW14} we presented numerical evidence for an
emergent approximate spontaneously broken $SO(2)$ symmetry with an associated
pseudo-Goldstone boson. Since the Goldstone boson has a small mass, it does
not qualify as a dual photon and the theory remains confining before one
reaches the RK point. While the exact diagonalization study alone is not
sufficient to come to this conclusion, it is fully consistent with it.

\begin{figure}[htb]
\centering
\includegraphics[scale=1.3]{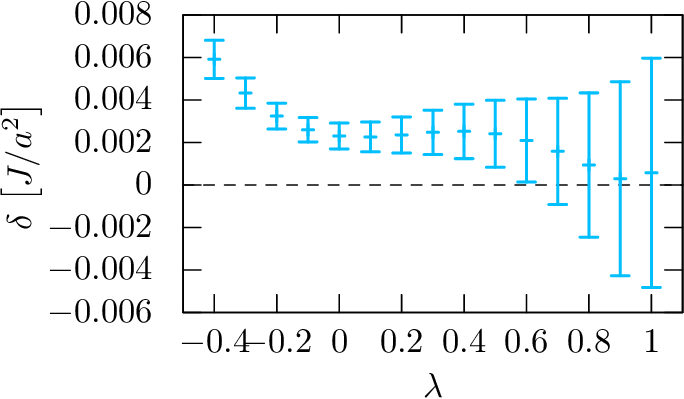} \\
\includegraphics[scale=1.3]{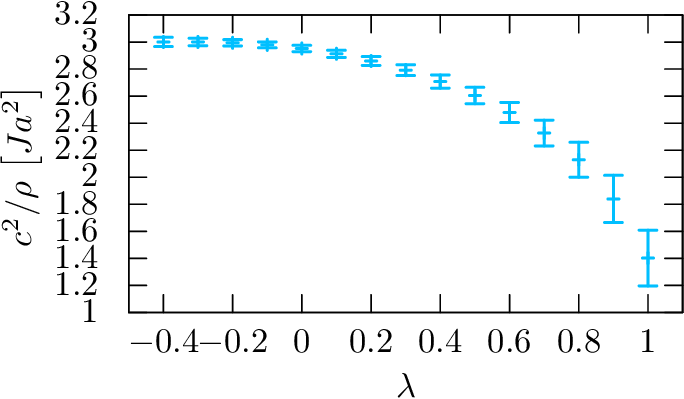}
\caption{[Color online] Results for the parameter $\delta$ (top) and the 
combination $\frac{c^2}{\rho}$ (bottom) of the effective theory from a fit to 
the exact diagonalization data for different $\lambda$ for the lattice sizes
$6\times 6$ and $8\times8$. The error bar that appears in the figures for each
value of $\lambda$ is the uncertainty of the calculated quantity from the fit.}
\label{qdm-deltafit}
\end{figure}

Using the analytic results of the effective theory obtained in the previous
section, we now estimate some low-energy parameters by comparison with the
exact diagonalization results for the rotor spectrum. Fig.~\ref{qdm-deltafit}
shows the results for the symmetry breaking parameter $\delta$ (top) and the
combination $\frac{c^2}{\rho}$ (bottom). These results have been obtained
from a global fit using data from $6\times 6$ and $8 \times 8$ lattices for
different values of $\lambda$. Note that $\frac{c^2}{\rho}$ is positive for
all values of $\lambda$, while  $\delta \geq 0$ approaches zero near the RK
point. Remarkably, the fit works rather well up to values of
$\lambda \approx 0.6$. Even though the errors are increasing near the RK
point, the results are still consistent with positive values of $\delta$, thus
indicating the absence of a phase transition before the RK point. This
suggests that the columnar phase extends all the way up to $\lambda=1$.
However, the precision reachable with the moderate volumes accessible to exact
diagonalization is not sufficient to definitively settle this issue. In
\citep{BBHJWW14} we have provided numerical evidence based on Monte Carlo data
obtained on much larger systems, which implies that $\delta$ remains positive
until one reaches the RK point, thus excluding a transition into the plaquette
phase.

One may wonder whether a mixed phase, sharing features of both the columnar
and the plaquette phase, would give rise to a similar finite-volume spectrum.
As we have pointed out in the previous section, in the mixed phase eight
vacuum states, separated by exponentially small energy gaps, are almost
degenerate in a  finite volume. This is qualitatively different from the rotor
spectrum observed in our exact diagonalization studies. First of all, the
energy of the rotor  states decreases inversely proportional to and not
exponentially with the volume. In addition, on the moderate volumes accessible
to exact diagonalization, the observed spectrum contains at least nine rotor
states, while the mixed phase would be characterized by eight low-energy
states separated from the rest of the spectrum by a gap.

Finally, Fig.~\ref{statecrossing} shows two sets of states with energies $E_1'$
and $E_2'$. These states have the quantum numbers ($CT_x,CT_y$) = (+,+), and
represent strings of non-zero electric flux
\begin{equation}
{\cal E}_i = \frac{1}{L_i} \sum_x E_{x,x+\hat i},
\end{equation}
wrapping around the periodic spatial volume. There are four states with energy
$E'_1$ with electric fluxes $({\cal E}_1,{\cal E}_2) = (\pm 1,0), 
(0,\pm 1)$, while there are two states with energy $E'_2$ with electric fluxes 
$(\pm 2,0), (0,\pm 2)$. The energy gaps of these states vanish at the RK point
$\lambda=1$. This implies that  at this point, flux strings cost zero energy 
thus signaling deconfinement and the spontaneous breakdown of the $U(1)$
center symmetry.

\section{Monte Carlo Results}
\label{MC}

Green's function Monte Carlo simulations have been applied earlier to the
square lattice quantum dimer model \citep{Syl05,Syl06,Ral08}, with lattice
sizes $L^2$ up to $L=48$. In our previous study \citep{BBHJWW14} we have used
a more efficient Monte Carlo algorithm that enabled us to reach volumes up to
$L^2 = 144 \times 144$ and temperatures down to $T = J/500$. Our algorithm is
based on the height variable representation of the quantum dimer model. More
details about the algorithm have been presented in \citep{BBHJWW14}.

Some Monte Carlo data, in particular those which provide convincing numerical
evidence that the columnar phase is realized in the square lattice quantum 
dimer model all the way to the RK point, are already shown in \citep{BBHJWW14}. 
Here we present new results and we give a more detailed explanation of the
results obtained earlier.

\begin{figure}
\begin{center}
\includegraphics[width=8cm]{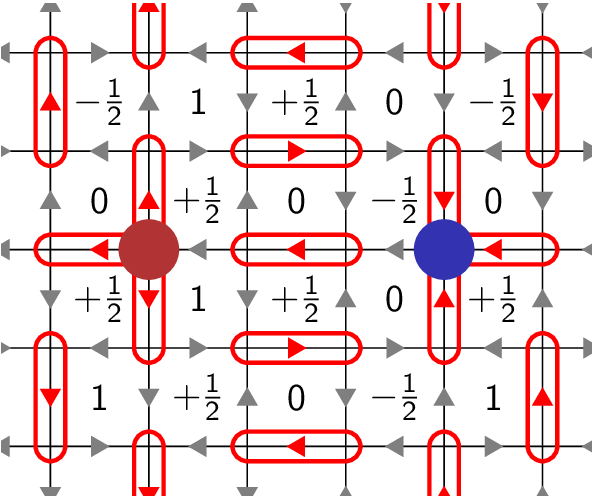}
\end{center}
\caption{[Color online] The presence of two external static charges violates
the dimer covering constraint \cite{BBHJWW14}.}
\label{externalCharges}
\end{figure}

\begin{figure}[htbp]
\begin{center}
\vbox{
\includegraphics[width=7cm]{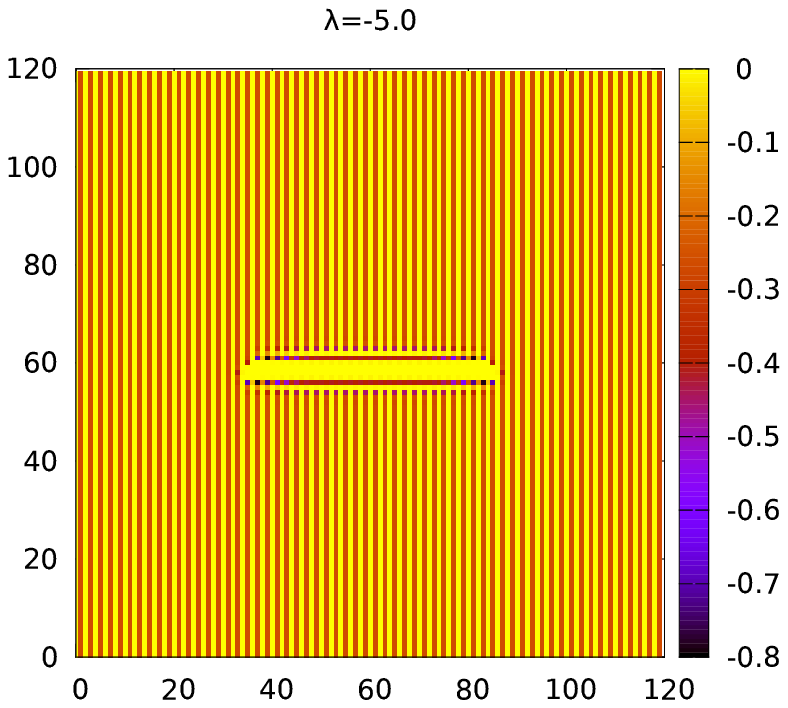}\\
\includegraphics[width=7cm]{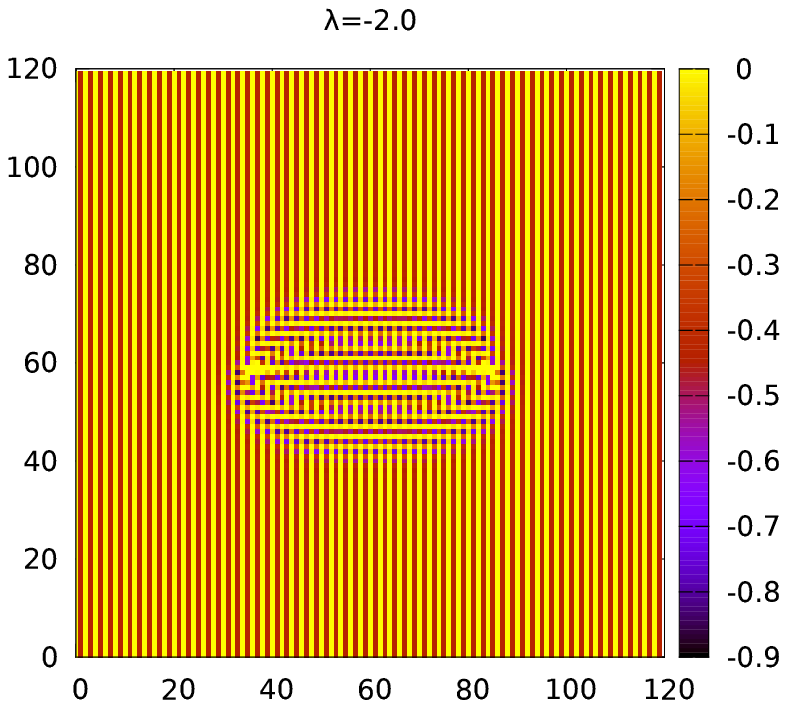}\\
\includegraphics[width=7cm]{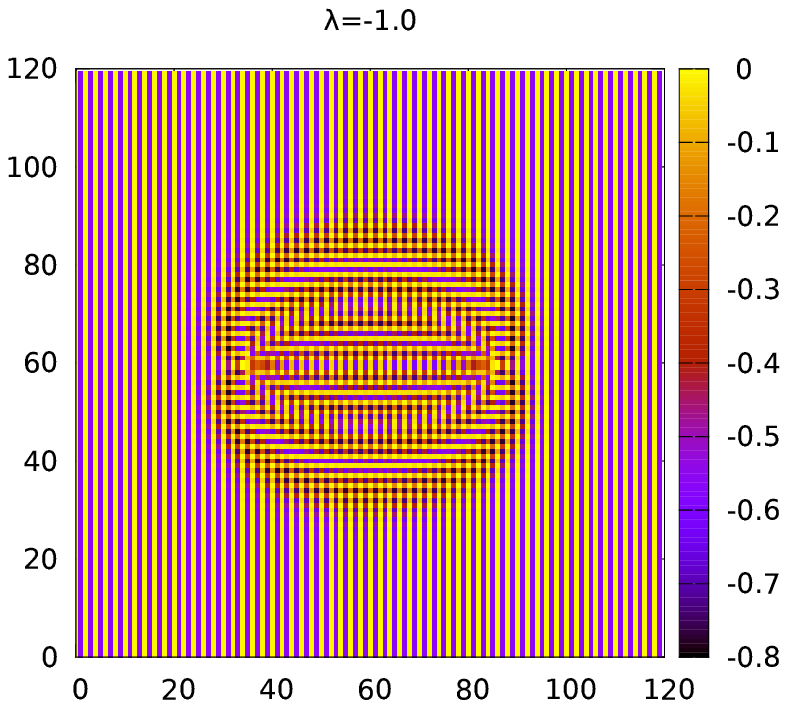}}
\end{center}
\caption{[Color online] (Top to bottom) Energy density $-J \langle U_\Box
+ U_\Box^\dagger\rangle$ on a $120 \times 120$ lattice in the presence of two
charges $\pm 2$, separated by 49 lattice spacings for $\beta J = 64$ and 
$\lambda= -5, -2, -1$.}
\label{energydensitym5m2m1}
\end{figure}

\begin{figure}[htbp]
\begin{center}
\includegraphics[width=10cm]{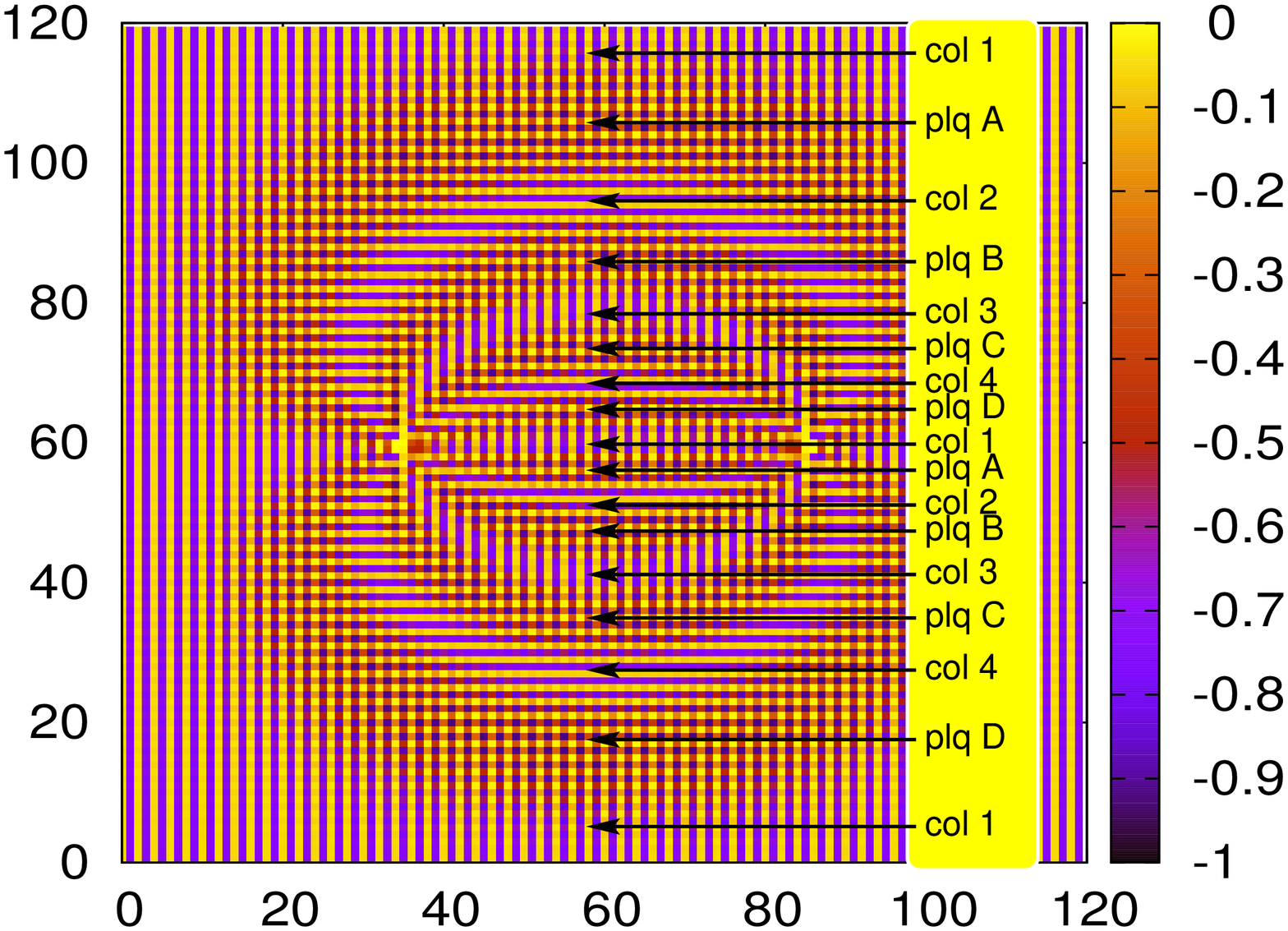}
\end{center}
\caption{[Color online] The appearance of plaquette order, which results from 
the interface between different columnar phases, in the eight strands. The 
result is obtained on a $120 \times 120$ lattice with $\beta J = 64$ and 
$\lambda=-0.5$.
}
\label{columnarplaquetteinterface}
\end{figure}

By putting two external static charges $\pm 2$ (relative to the staggered
charge background) into the system, one violates the dimer covering
constraint. As depicted in Fig.~\ref{externalCharges}, this leads to two 
defects, associated with three dimers that overlap at the same lattice point. 
The two static charges, separated by an odd number of lattice spacings, are 
connected by an electric flux string and are thus confined. In addition, the 
flux string fractionalizes into eight individual strands --- displaying the
four plaquette phases --- which each carry electric flux $\frac{1}{4}$, thus
adding up to the total flux 2. The energy density
$-J \langle U_\Box + U_\Box^\dagger\rangle$ for $\lambda = -5, -2, -1$ is
shown in Fig.~\ref{energydensitym5m2m1}: one notices
that, as one moves from large negative values of $\lambda$ towards
$\lambda \approx 0$, the individual strands emerge around $\lambda \approx -2$. 
Furthermore, inside the different strands plaquette order is present. These 
regions of plaquette order are interfaces separating the various columnar
phases. In fact, Fig.~\ref{columnarplaquetteinterface} implies that as one
moves from bottom to top, each of the four possible columnar phases is visited
once. The same is true for the four degenerate plaquette  orders.

\begin{figure}%[htbp]
\begin{center}
\includegraphics[width=10cm]{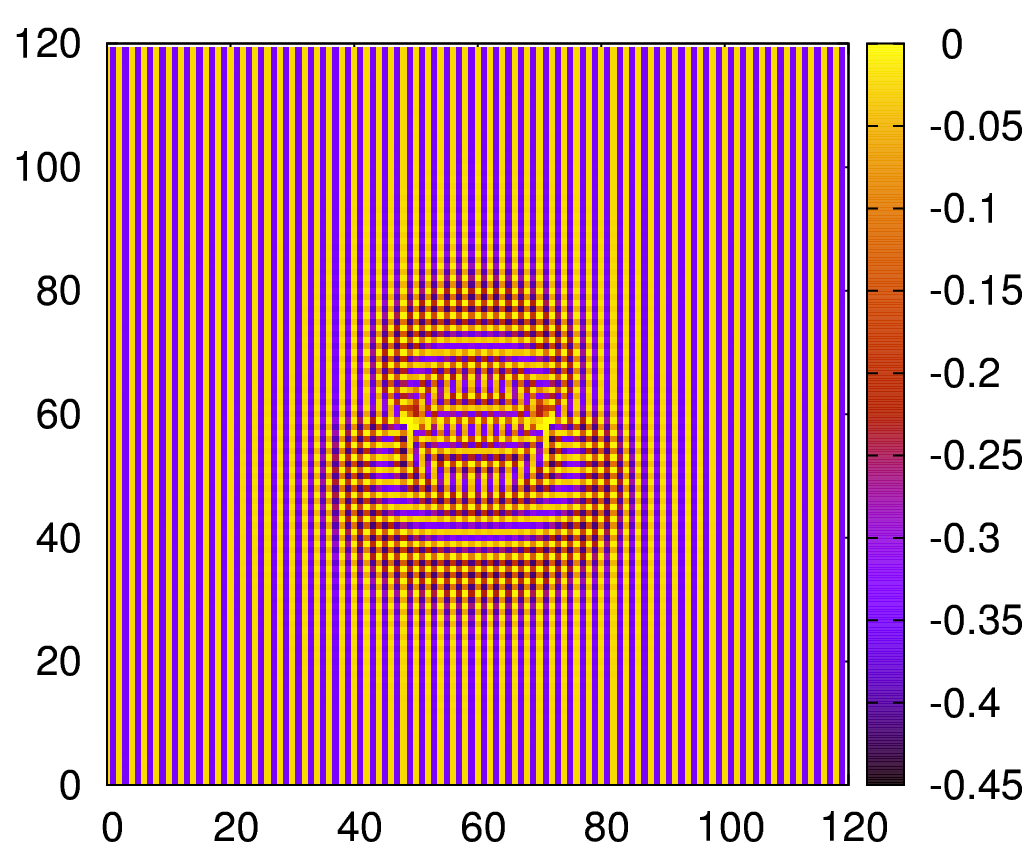}
\end{center}
\caption{[Color online] Energy density 
$-J \langle U_\Box + U_\Box^\dagger\rangle$ on a $120 \times 120$ lattice in the 
presence of two charges $\pm 2$, separated by 43 lattice spacings for 
$\beta J = 64$ and $\lambda = -0.5$.}
\label{separation43}
\end{figure}

\begin{figure}%[htbp]
\begin{center}
\includegraphics[width=10cm]{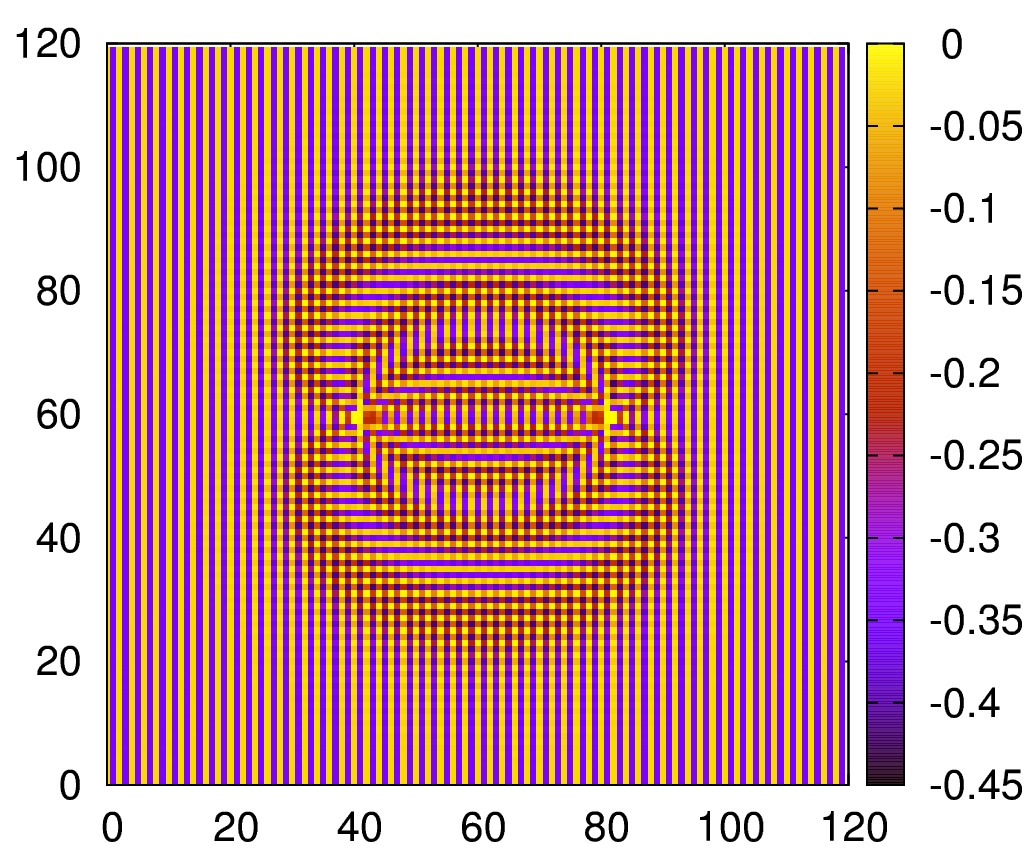}
\end{center}
\caption{[Color online] Energy density 
$-J \langle U_\Box + U_\Box^\dagger\rangle$ on a $120 \times 120$ lattice in the 
presence of two charges $\pm 2$, separated by 53 lattice spacings for 
$\beta J = 64$ and $\lambda=-0.5$.}

\label{separation53}
\end{figure}

In the presence of two external charges $\pm 2$ separated along the $x$-axis,
both translation and rotation invariance are explicitly broken, while the
reflection on the $x$-axis remains an exact symmetry. As a result, one of the
columnar phases, with the columns oriented in the $y$-direction, is
energetically favored. Interestingly, Fig.~\ref{separation43} shows that, for
$\lambda=-0.5$ an asymmetric distribution of the eight strands is observed,
thus indicating the spontaneous breakdown of the reflection symmetry. Strictly
speaking, in an infinite volume spontaneous breaking of the reflection symmetry
 only arises when the distance between the charges also approaches infinity. 
At finite distances, the two
asymmetric flux  patterns, which are related to one another by reflection,
coexist with each other through quantum tunneling. As the charges are
separated further and further, still assuming an infinite volume, 
tunneling is exponentially suppressed. When we consider a finite volume, the
asymmetry in the flux distribution disappears when the distance between the
charges becomes compatible with the lattice size (Fig.~\ref{separation53}).
Squeezing the flux distribution into a small volume leads to a restoration
 of the spontaneously broken reflection symmetry
due to finite-size effects. This scenario also arises for other negative
values of $\lambda$.

\section{Conclusions}
\label{conclusions}

We have investigated the finite-volume energy spectrum of the square lattice
quantum dimer model using both exact diagonalization and quantum Monte Carlo.
By comparison with analytic predictions we have consolidated our previous
evidence that the columnar phase extends all the way to the RK point, without
any intervening plaquette or mixed phases. In addition, we have studied a soft
pseudo-Goldstone mode that becomes massless at the RK point but still dominates
a large region in parameter space away from it. This mode is described by a 
systematic low-energy effective field theory whose parameters we have extracted
by comparison of numerical data with analytic predictions of the effective
theory. It will be an interesting topic for future studies to investigate the
possible role of the soft mode for the preformation of pairs in the pseudogap
regime of high-temperature superconductors. This could be done in the context
of hole-doped quantum dimer models 
\citep{Bal05,Poi06,Ral07,Pap07,Rib07,Poi08,Lam12,Lam13}.

We have also studied the internal structure of the strings connecting external
charges embedded in the confining columnar phase. For topological reasons, the
string fractionalizes into strands, each carrying electric flux $\frac{1}{4}$.
The flux strands play the role of interfaces separating the four realizations
of the columnar phase. As we noted earlier \citep{BBHJWW14}, the interior of
the flux strands shows plaquette order, despite the fact that the plaquette
phase is not stable in the bulk. The interfaces that separate two columnar
phases 1 and 3 or 2 and 4, with the columns oriented in the same direction,
show the universal phenomenon of complete wetting. This manifests itself by
the appearance of a third columnar phase at the interface, with its columns
oriented in an orthogonal direction. Hence, a 1-3 interface splits into two
1-2-3 or 1-4-3 interfaces. Remarkably, reflections on the lattice axis 
connecting charges $\pm 2$ are spontaneously broken, which gives rise to
asymmetric electric flux profiles.

As we have seen, the simple square lattice quantum dimer model has a rich 
confining dynamics, characterized by strings with an intriguing anatomy. 
Understanding these dynamics required the interplay between numerical 
simulations and analytic effective field theory calculations. It will be 
interesting and promising to apply this strategy to quantum dimer models with
other lattice geometries \citep{SBMVM15,SMB15}.

\section*{Acknowledgments}

DB acknowledges interesting discussions with A.\ L\"auchli. CPH thanks the
members of the ITP at Bern University for their hospitality, and acknowledges
support through the project {\it Redes Tem\'aticas de Colaboraci\'on
Acad\'emica 2013}, UCOL-CA-56. The research leading to these results has
received funding from the Schweizerischer Na\-tio\-nal\-fonds and from the
European Research Council under the European Union's Seventh Framework
Programme (FP7/2007-2013)/ ERC grant agreement 339220.

\begin{appendix}

\section{Symmetries and Candidate Phases}
\label{appendixA}

It is important to know how the four order parameters transform under the
symmetries $CT_x, CT_y, O, CO', R_x$, and $R_y$ of the square lattice
quantum dimer model. This is illustrated in Table \ref{Osymmetries}.

\begin{table}[h]
\begin{center}
\begin{tabular}{|c|c|c|c|c|c|c|}
\hline 
$S$ & CT$_x$ & CT$_y$ & O & CO' & R$_x$ & R$_y$ \\
\hline
\hline
$M_{11}[^S{\cal C}]$&-$M_{11}[{\cal C}]$&-$M_{22}[{\cal C}]$& $M_{21}[{\cal C}]$&
                    -$M_{21}[{\cal C}]$&-$M_{22}[{\cal C}]$&-$M_{11}[{\cal C}]$ \\
\hline
$M_{12}[^S{\cal C}]$& $M_{21}[{\cal C}]$&-$M_{12}[{\cal C}]$&-$M_{11}[{\cal C}]$&
                    -$M_{22}[{\cal C}]$&-$M_{12}[{\cal C}]$& $M_{21}[{\cal C}]$ \\
\hline
$M_{21}[^S{\cal C}]$& $M_{12}[{\cal C}]$& $M_{21}[{\cal C}]$&-$M_{22}[{\cal C}]$&
                    -$M_{11}[{\cal C}]$& $M_{21}[{\cal C}]$& $M_{12}[{\cal C}]$ \\
\hline
$M_{22}[^S{\cal C}]$& $M_{22}[{\cal C}]$&-$M_{11}[{\cal C}]$&-$M_{12}[{\cal C}]$&
                    -$M_{12}[{\cal C}]$&-$M_{11}[{\cal C}]$& $M_{22}[{\cal C}]$ \\
\hline
\end{tabular}
\end{center}
\caption{Transformation properties of the order parameters $M_{ij}$ under
the symmetries  $S =$  CT$_x$, CT$_y$, O, CO', R$_x$, R$_y$ . The order
parameter $M_{ij}[^S{\cal C}]$ evaluated in the transformed configuration
$^S{\cal C}$ as a function of the order parameters $M_{ij}[{\cal C}]$ evaluated
in the original configuration ${\cal C}$.}
\label{Osymmetries}
\end{table}

\begin{table}[h]
\begin{center}
\begin{tabular}{|c|c|c|c|c|c|c|}
\hline 
$S$ & CT$_x$ & CT$_y$ & O & CO' & R$_x$ & R$_y$ \\
\hline
\hline
$^S$1 & 1 & 3 & 2 & 4 & 3 & 1 \\
\hline
$^S$2 & 4 & 2 & 3 & 3 & 2 & 4 \\
\hline
$^S$3 & 3 & 1 & 4 & 2 & 1 & 3 \\
\hline
$^S$4 & 2 & 4 & 1 & 1 & 4 & 2 \\
\hline
\end{tabular}
\end{center}
\caption{Transformation properties of the four columnar phases 1, 2, 3, 4
under the symmetries $S =$  CT$_x$, CT$_y$, O, CO', R$_x$, R$_y$.}
\label{Csymmetries}
\end{table}

\begin{table}[h]
\begin{center}
\begin{tabular}{|c|c|c|c|c|c|c|}
\hline 
$S$ & CT$_x$ & CT$_y$ & O & CO' & R$_x$ & R$_y$ \\
\hline
\hline
$^S$A & D & B & B & C & B & D \\
\hline
$^S$B & C & A & C & B & A & C \\
\hline
$^S$C & B & D & D & A & D & B \\
\hline
$^S$D & A & C & A & D & C & A \\
\hline
\end{tabular}
\end{center}
\caption{Transformation properties of the four plaquette phases A, B, C, D
under the symmetries $S =$  CT$_x$, CT$_y$, O, CO', R$_x$, R$_y$.}
\label{Psymmetries}
\end{table}

\begin{table}[h]
\vskip0.5cm
\begin{center}
\begin{tabular}{|c|c|c|c|c|c|c|}
\hline 
$S$ & CT$_x$ & CT$_y$ & O & CO' & R$_x$ & R$_y$ \\
\hline
\hline
$^S$A1 & D1 & B3 & B2 & C4 & B3 & D1 \\
\hline
$^S$A2 & D4 & B2 & B3 & C3 & B2 & D4 \\
\hline

$^S$B2 & C4 & A2 & C3 & B3 & A2 & C4 \\
\hline
$^S$B3 & C3 & A1 & C4 & B2 & A1 & C3 \\
\hline
$^S$C3 & B3 & D1 & D4 & A2 & D1 & B3 \\
\hline
$^S$C4 & B2 & D4 & D1 & A1 & D4 & B2 \\
\hline
$^S$D4 & A2 & C4 & A1 & D1 & C4 & A2 \\
\hline
$^S$D1 & A1 & C3 & A2 & D4 & C3 & A1 \\
\hline
\end{tabular}
\end{center}
\caption{Transformation properties of the eight mixed phases A1, A2, B2,
B3, C3, C4, D4, D1 under the symmetries $S =$  CT$_x$, CT$_y$, O, CO', R$_x$,
R$_y$.}
\label{Msymmetries}
\end{table}

In tables \ref{Csymmetries}, \ref{Psymmetries}, and \ref{Msymmetries}, we show
how the different symmetries $CT_x, CT_y, O, CO', R_x, R_y$ act on the
columnar, plaquette, and mixed phases, respectively.

\end{appendix}


\begin{thebibliography}{10}

\bibitem{BM86}
J.\ C.\ Bednorz and K.\ A.\ M\"uller, Z.\ Phys.\ B: Condens.\ Matter
\textbf{64}, 189 (1986).

\bibitem[Rokhsar and Kivelson (1988)]{Rok88}
D.\ S.\ Rokhsar and S.\ A.\ Kivelson, Phys.\ Rev.\ Lett.\ \textbf{61}, 2376
(1988).

\bibitem{And87}
P.\ W.\ Anderson, Science \textbf{235}, 1196 (1987).

\bibitem{Sac89}
S.\ Sachdev, Phys.\ Rev.\ B \textbf{40}, 5204 (1989).

\bibitem{Lev90}
% Equivalence of the dimer resonating-valence-bond problem to the quantum
% roughening problem, 
L.\ S.\ Levitov, Phys.\ Rev.\ Lett.\ \textbf{64}, 92 (1990).

\bibitem{Leu96}
% Columnar dimer and plaquette resonating valence bond states of the quantum
% dimer model,
P.\ W.\ Leung, K.\ C.\ Chiu, and K.\ J.\ Runge, Phys.\ Rev.\ B \textbf{54},
12938 (1996).

\bibitem{Moe02}
R.\ Moessner, S.\ L.\ Sondhi, and E.\ Fradkin, Phys.\ Rev.\ B \textbf{65},
024504 (2002). 

\bibitem{Hen04}
% From classical to quantum dynamics at Rokhsar-Kivelson points,
C.\ L.\ Henley, J.\ Phys.: Condens.\ Matter \textbf{16}, S891 (2004).

\bibitem{Ale05}
% Interacting classical dimers on the square lattice,
F.\ Alet, J.\ L.\ Jacobsen, G.\ Misguich, V.\ Pasquier, F.\ Mila, and 
M.\ Troyer, Phys.\ Rev.\ Lett.\ \textbf{94}, 235702 (2005).

\bibitem{Ale06}
% Classical dimers with aligning interactions on the square lattice,
F.\ Alet, Y.\ Ikhlef, J.\ L.\ Jacobsen, G.\ Misguich, and V.\ Pasquier,
Phys.\ Rev.\ E \textbf{74}, 041124 (2006).

\bibitem{Cha10}
% Phase diagram of an extended classical dimer model,
D.\ Charrier and F.\ Alet, Phys.\ Rev.\ B \textbf{82}, 014429 (2010).

\bibitem{Can10}
% Spin Hamiltonians with resonating-valence-bond ground states,
J.\ Cano and P.\ Fendley, Phys.\ Rev.\ Lett.\ \textbf{105}, 067205 (2010). 

\bibitem{Alb10}
% Critical correlations for short-range valence-bond wave functions on the
% square lattice,
A.\ F.\ Albuquerque and F.\ Alet, Phys.\ Rev.\ B \textbf{82}, 180408(R)
(2010).


\bibitem{Tan11}
% Properties of resonating-valence-bond spin liquids and critical dimer
% models,
Y.\ Tang, A.\ W.\ Sandvik, and C.\ L.\ Henley, Phys.\ Rev.\ B \textbf{84},
174427 (2011).

\bibitem{Lam13}
% Statistics of holes and nature of superfluid phases in quantum dimer models,
C.\ A.\ Lamas, A.\ Ralko, M.\ Oshikawa, D.\ Poilblanc, and P.\ Pujol,
Phys.\ Rev.\ B \textbf{87}, 104512 (2013).

\bibitem{Syl05}
% Continuous-time diffusion Monte Carlo and the quantum dimer model,
O.\ F.\ Syljuasen, Phys.\ Rev.\ B \textbf{71}, 020401(R) (2005).

\bibitem{Syl06}
% The plaquette phase of the square lattice quantum dimer model,
O.\ F.\ Syljuasen, Phys.\ Rev.\ B \textbf{73}, 245105 (2006).

\bibitem{Ral08}
A.\ Ralko, D.\ Poilblanc, and R.\ Moessner, Phys.\ Rev.\ Lett.\ \textbf{100},
037201 (2008).

% NEW reference
\bibitem{BBHJWW14}
D.\ Banerjee, M.\ B\"ogli, C.\ P.\ Hofmann, F.-J.\ Jiang, P.\ Widmer, and
U.-J.\ Wiese, Phys.\ Rev.\  B \textbf{90}, 245143 (2014).

\bibitem{Hor81}
D.\ Horn, Phys.\ Lett.\ \textbf{100B}, 149 (1981).

\bibitem{Orl90}
P.\ Orland and D.\ Rohrlich, Nucl.\ Phys.\ B \textbf{338}, 647 (1990).

\bibitem{Cha97}
S.\ Chandrasekharan and U.-J.\ Wiese, Nucl.\ Phys.\ B \textbf{492}, 455
(1997).

\bibitem{tHo79}
G.\ 't Hooft, Nuci.\ Phys.\ B \textbf{153}, 141 (1979). 

\bibitem[Moessner and Raman (2008)]{MR08}
R.\ Moessner and K.\ S.\ Raman, arXiv:0809.3051.

% NEW reference
\bibitem[Moessner and Sondi (2001)]{MS01}
% Resonating valence bond phase in the triangular lattice quantum dimer model
R.\ Moessner and S.\ L.\ Sondhi, Phys.\ Rev.\ Lett. \textbf{86}, 1881 (2001). 

% NEW reference
\bibitem[Ivanov (2004)]{Iva04}
% Vortexlike elementary excitations in the Rokhsar-Kivelson dimer model on the
% triangular lattice
D.\ A.\ Ivanov, Phys.\ Rev.\ B \textbf{70}, 094430 (2004).

% NEW reference
\bibitem[Ralko et al.(2005)]{RFBIM05}
% Zero-temperature properties of the quantum dimer model on the triangular
% lattice
A.\ Ralko, M.\ Ferrero, F.\ Becca, D.\ Ivanov, and F.\ Mila, Phys.\ Rev.\ B
\textbf{71}, 224109 (2005).

% NEW reference
\bibitem[Huse et al (2003)]{HKMS03}
% Coulomb and liquid dimer models in three dimensions
D.\ A.\ Huse, W.\ Kraut, R.\ Moessner, and S.\ L.\ Sondhi, Phys.\ Rev.\ Lett.\
\textbf{91}, 167004 (2003).

% NEW reference
\bibitem[Moessner and Sondi (2003)]{MS03}

% Three-dimensional resonating-valence-bond liquids and their excitations
R.\ Moessner and S.\ L.\ Sondhi, Phys.\ Rev.\ B \textbf{68}, 184512 (2003). 

\bibitem{Her04}
M.\ Hermele, M.\ P.\ A.\ Fisher, and L.\ Balents, Phys.\ Rev.\ B \textbf{69},
064404
(2004).

% NEW reference
\bibitem[Moessner et al. (2000)]{MSC00}
% Two-Dimensional Periodic Frustrated Ising Models in a Transverse Field
R.\ Moessner, S.\ L.\ Sondhi, and P.\ Chandra, Phys.\ Rev.\ Lett. \textbf{84},
4457 (2000). 

% NEW reference
\bibitem[Zeng and Elser (1995)]{ZE95}
% Quantum dimer calculations on the spin-1/2 Kagome Heisenberg antiferromagnet
C.\ Zeng and V.\ Elser, Phys.\ Rev.\ B \textbf{51}, 8318 (1995).

% NEW reference
\bibitem[Nikolic and Senthil (2003)]{NS03}
% Physics of low-energy singlet states of the Kagome lattice quantum
% Heisenberg antiferromagnet
P.\ Nikolic and T.\ Senthil, Phys.\ Rev.\ B \textbf{68}, 214415 (2003). 

% NEW reference
\bibitem[Moessner et al. (2001)]{MSC01}
% Phase diagram of the hexagonal lattice quantum dimer model
R.\ Moessner, S.\ L.\ Sondhi, and P.\ Chandra, Phys.\ Rev.\ B \textbf{64},
144416 (2001). 

\bibitem[Antonov (1907)]{Ant07}
G.\ Antonov, J.\ Chem.\ Phys.\ \textbf{5}, 372 (1907).

% Putting competing orders in their place near the Mott transition. II. The
%doped quantum dimer model
\bibitem{Bal05}
L.\ Balents, L.\ Bartosch, A.\ Burkov, S.\ Sachdev, and K.\ Sengupta,
Phys.\  Rev.\ B \textbf{71}, 144509 (2005).

% Doping quantum dimer models on a square lattice
\bibitem{Poi06}
D.\ Poilblanc, F.\ Alet, F.\ Becca, A.\ Ralko, F.\ Trousselet, and F.\ Mila,
 Phys.\ Rev.\ B \textbf{74}, 014437 (2006).

% Phase separation and flux quantization in the doped quantum dimer model on
%square and triangular lattices
\bibitem{Ral07}
A.\ Ralko, F.\ Mila, and D.\ Poilblanc,
Phys.\ Rev.\ Lett. \textbf{99}, 127202  (2007).

% Quantum criticality, lines of fixed points, and phase separation in doped
%two-dimensional quantum dimer models
\bibitem{Pap07}
S.\ Papanikolaou, E.\ Luijten, and E.\ Fradkin,
Phys.\ Rev.\ B \textbf{76},  134514 (2007).

% Single hole and vortex excitations in the doped Rokhsar-Kivelson quantum
%dimer model on the triangular lattice
\bibitem{Rib07}
H.\ Ribeiro, S.\ Bieri, and D.\ Ivanov,
Phys.\ Rev.\ B \textbf{76}, 172301  (2007).

% Properties of holons in the quantum dimer model
\bibitem{Poi08}
D.\ Poilblanc,
Phys.\ Rev.\ Lett.\ \textbf{100}, 157206 (2008).

% Statistical transmutation in doped quantum dimer models
\bibitem{Lam12}
C.\ A.\ Lamas, A.\ Ralko, D.\ C.\ Cabra, D.\ Poilblanc, and P.\ Pujol, Phys.\
Rev.\ Lett.\textbf{109}, 016403 (2012).

\bibitem[Schlittler et al.(2015)]{SBMVM15}
T.\ M.\ Schlittler, T.\ Barthel, G.\ Misguich, J.\ Vidal, and R.\ Mosseri,
%Phase diagram of an extended quantum dimer model on the hexagonal lattice
arXiv:1507.04643.

%\bibitem[Giuliani and Lieb (2015)]{GL15}
%A.\ Giuliani and E.\ H.\ Lieb,
%Columnar phase in quantum dimer models
%Phys.\ A: Math.\ Theor.\ , \textbf{48}, 235203 (2015).

\bibitem[Schlittler et al.(2015)]{SMB15}
T.\ M.\ Schlittler, R.\ Mosseri, and T.\ Barthel,
%Phase diagram of the hexagonal lattice quantum dimer model: order parameters,
%ground-state energy, and gaps
arXiv:1501.02242.




\end{thebibliography}
\end{document}